\documentclass[fleqn,usenatbib]{mnras}

\usepackage{newtxtext,newtxmath}

\usepackage[T1]{fontenc}

\DeclareRobustCommand{\VAN}[3]{#2}
\let\VANthebibliography\thebibliography
\def\thebibliography{\DeclareRobustCommand{\VAN}[3]{##3}\VANthebibliography}

\usepackage{graphicx}	
\usepackage{amsmath}	
\usepackage{amssymb}	
\usepackage{booktabs}
\usepackage{xcolor}
\usepackage{ulem}

\newcommand{\rcl}{\ifmmode{r_{\mathrm{cl}}}\else$r_{\mathrm{cl}}$\fi}
\newcommand{\Tcl}{\ifmmode{T_{\mathrm{cl}}}\else$T_{\mathrm{cl}}$\fi}
\newcommand{\ncl}{\ifmmode{n_{\mathrm{cl}}}\else$n_{\mathrm{cl}}$\fi}
\newcommand{\Tmix}{\ifmmode{T_{\mathrm{mix}}}\else$T_{\mathrm{mix}}$\fi}
\newcommand{\Twind}{\ifmmode{T_{\mathrm{wind}}}\else$T_{\mathrm{wind}}$\fi}
\newcommand{\Tdest}{\ifmmode{T_{\mathrm{dest}}}\else$T_{\mathrm{dest}}$\fi}
\newcommand{\tcc}{\ifmmode{t_{\mathrm{cc}}}\else$t_{\mathrm{cc}}$\fi}
\newcommand{\dcell}{\ifmmode{d_{\mathrm{cell}}}\else$d_{\mathrm{cell}}$\fi}

\newcommand{\tminmix}{\ifmmode{t_{\mathrm{cool,minmix}}}\else$t_{\mathrm{cool,minmix}}$\fi}
\newcommand{\tmax}{\ifmmode{t_{\mathrm{cool,max}}}\else$t_{\mathrm{cool,max}}$\fi}
\newcommand{\tmixCC}{\ifmmode{t_{\mathrm{dest,mix}}}\else$t_{\mathrm{dest,mix}}$\fi}
\newcommand{\tmixCool}{\ifmmode{t_{\mathrm{cool,mix}}}\else$t_{\mathrm{cool,mix}}$\fi}

\newcommand{\thot}{$t_{\mathrm{cool,wind}}$}
\newcommand{\lrg}{\texttt{LRG}}
\newcommand{\med}{\texttt{MED}}
\newcommand{\sml}{\texttt{SML}}

\newcommand{\cc}{cm$^{-3}$}
\newcommand{\Msun}{M$_{\odot}$}

\newcommand{\Mach}{\ifmmode{\mathcal{M}}\else$\mathcal{M}$\fi}

\newcommand{\fourEfive}{$4 \times 10^{5}$}
\newcommand{\fourEfour}{$4 \times 10^{4}$}

\newcommand{\eightEThree}{$8 \times 10^{3}$}

\newcommand{\tenFour}{$10^{4}$}
\newcommand{\tenThree}{$10^{3}$}

\providecommand{\p}[2]{\ensuremath{\frac{\partial {#1}}{\partial {#2}}}}
\newcommand{\bnabla}{{\mbox{\boldmath$\nabla$}}}

\title[Dusty Clouds in Hot Winds]{The Survival of Multiphase Dusty Clouds in Hot Winds}

\author[Farber \& Gronke]{
Ryan J. Farber,$^{1}$\thanks{E-mail: rjfarber@umich.edu}
and Max Gronke,$^{2,3}$\thanks{Hubble fellow}
\\
$^{1}$Department of Astronomy, University of Michigan, 1085 S. University Ave., Ann Arbor, MI 48109, USA\\
$^{2}$Department of Physics \& Astronomy, Johns Hopkins University, 
    Bloomberg Center, 3400 N. Charles St., Baltimore, MD 21218, USA\\
${}^3$    Max Planck Institut fur Astrophysik, Karl-Schwarzschild-Straße 1, D-85748 Garching bei München, Germany\\
}

\date{Accepted XXX. Received YYY; in original form ZZZ}

\pubyear{2021}

\begin{document}
\label{firstpage}
\pagerange{\pageref{firstpage}--\pageref{lastpage}}
\maketitle

\begin{abstract}
Much progress has been made recently in the acceleration of $\sim10^{4}$\,K clouds to explain absorption-line measurements of the circumgalactic medium and the warm, atomic phase of galactic winds. However, the origin of the cold, molecular phase in galactic winds has received relatively little theoretical attention. Studies of the survival of $\sim10^{4}$\,K clouds suggest efficient radiative cooling may enable the survival of expelled material from galactic disks. Alternatively, gas colder than 10$^4$\,K may form within the outflow, including molecules if dust survives the acceleration process. We explore the survival of dusty clouds in a hot wind with three-dimensional hydrodynamic simulations including radiative cooling and dust modeled as tracer particles. We find that cold $\sim10^{3}$\,K gas can be destroyed, survive, or transformed entirely to $\sim 10^4\,$K gas. We establish analytic criteria distinguishing these three outcomes which compare characteristic cooling times to the system's `cloud crushing' time. In contrast to typically studied $\sim10^{4}$\,K clouds, colder clouds are entrained faster than the drag time as a result of efficient mixing. We find that while dust can in principle survive embedded in the accelerated clouds, the survival fraction depends critically on the time dust spends in the hot phase and on the effective threshold temperature for destruction. We discuss our results in the context of polluting the circumgalactic medium with dust and metals, as well as understanding observations suggesting rapid acceleration of molecular galactic winds and ram pressure stripped tails of jellyfish galaxies.
\end{abstract}

\begin{keywords}
galaxies:evolution -- dust -- ISM: clouds -- hydrodynamics -- Galaxy: halo -- ISM: molecules
\end{keywords}


\section{Introduction}
\label{sec:intro}

Since the historic discovery of massive outflowing filaments from M82 \citep{Lynds1963}, galactic outflows have been observed ubiquitously \citep{Veilleux2005,Veilleux2020} both locally (\citealt{Lehnert1996}; \citealt{Martin2005}; \citealt{Rupke2005}) and at high redshifts \citep[e.g.,][]{Shapley2003}. Outflows are crucial in regulating the mass-metallicity relationship in galaxies (\citealt{Larson1974}), polluting the circumgalactic medium (CGM) and intergalactic medium (IGM) with metals (\citealt{Mac1999}; \citealt{Steidel2010}; \citealt{Booth2012}). Moreover, galactic outflows may help to address one of the most pressing puzzles in galaxy formation: the missing baryons problem (\citealt{Bell2003}).

While the amount of ``baryons'' (gas and stars) in galaxy clusters compared to their total mass is roughly consistent with the cosmological baryon fraction observed by \textit{Planck} (\citealt{Collaboration2020}), a discrepancy arises at lower masses, as large as an order of magnitude for L$_{*}$ galaxies and several orders of magnitude for dwarfs (\citealt{Dai2010}). This discrepancy may be explained by galactic outflows either ejecting material from the galaxy or preventing accretion of cosmological gas onto the galaxy. While active galactic nuclei likely dominate the outflow energetics of more massive galaxies than the Milky Way (\citealt{Croton2006}), outflows in lower mass galaxies are consistent with the energy budget of stellar winds and supernovae (\citealt{Somerville2015}).

In the standard model of stellar feedback (\citealt{Chevalier1985}; CC85), supernovae shock-heat gas to $10^{6-7}\,$K, accelerating the gas to high velocities. X-ray observations of hot outflows find superb agreement between the mass outflow rate of hot gas with the CC85 model (\citealt{Strickland2007}), yet estimates of net mass loading are far lower than what is needed to regulate star formation (\citealt{Mac1999}).

At the same time, millimeter and submillimeter observations detect rapidly outflowing molecular gas around nearby starbursts e.g., NGC 253 (\citealt{Bolatto2013}), the Galaxy (\citealt{Di2020}; \citealt{su2021molecular}), and at high redshifts (z > 4, cf. \citealt{Spilker2020}). Bright CO features close to the launching radius suggest short dynamical time-scales, strongly pointing to entrainment of galactic gas (\citealt{Walter2017}).

Classically, the acceleration of a cold cloud in a hot wind is known as the cloud crushing problem (\citealt{Cowie1977}; \citealt{McKee1977}; \citealt{Balbus1982}; \citealt{Stone1992}; \citealt{Klein1994}; \citealt{Mac1994}; \citealt{Xu1995}). Since the cloud destruction time-scale due to hydrodynamic instabilities is shorter than the acceleration time-scale of (hydrodynamic) ram pressure from the wind acting on the cloud (cf. \citealt{Klein1994}), subsequent work has focused on the ability of magnetic fields (\citealt{McCourt2015}; \citealt{berlok2019}), conduction (\citealt{Armillotta2016}; \citealt{bruggen2016launching}; \citealt{cottle2020launching}) and the Mach cone from a supersonic flow (\citealt{Scannapieco2015}) to prolong the cloud lifetime. However, clouds are nevertheless eventually destroyed by the hot medium. 

Recent work (\citealt{Marinacci2010}; \citealt{banda2016filament}; \citealt{Armillotta2017}; \citealt{Gronke2018}; \citealt{Sparre2019}; \citealt{Sparre2020}; \citealt{Gronke2020Cloudy}; \citealt{Li2020}; \citealt{Kanjilal2021}; \citealt{Abruzzo2021}; \citealt{gronke2021survival}) finds that efficient radiative cooling can enable a purely hydrodynamic acceleration of cold clouds (and similarly for cold streams, \citealt{Mandelker2020}) in the case of a transonic wind, as occurs at the launching radius of a galactic wind. Initially, mixing reduces the cold mass (which we explain in this work); however clouds that survive can grow an order of magnitude in mass while simultaneously acquiring the momentum of the hot wind.


Alternatively, the cold gas may form \textit{in-situ} as the outflowing wind cools both adiabatically and with increased efficiency radiatively as the density of the wind is enhanced via  incorporation of destroyed clouds (\citealt{Wang1994}; \citealt{Thompson2016}). In this model, the cold phase is naturally expected to be co-moving with the hot phase (and growing in mass) for distances greater than the cooling radius\footnote{The cooling radius marks the point at which the wind has adiabatically cooled sufficiently that the radiative cooling time becomes shorter than the dynamical time (cf. equation 6 in \citealp{Thompson2016})}, a few hundred parsecs above the disk. However, at least some systems exhibit properties apparently incompatible with bulk cooling \citep{lochhaas2021characteristic}. Moreover, in the case of M82 where detailed multiwavelength measurements abound: (i) cold outflowing gas is observed to diminish in mass flux with distance above the disk (consistent with cloud crushing), and (ii) dust is observed to be spatially coincident with outflowing HI and CO gas suggesting survival at least of dust from the galactic disk (\citealt{Leroy2015}).

While atomic gas can plausibly form in the thermal instability model, molecules cannot so simply re-form from the ashes of the mixed gas. Efficient formation of molecules requires cold dust grains that are dynamically $\sim$stationary with respect to atomic gas\footnote{Since the kinetic energy of the grains must be less than the binding energy of atomic H on the grains.} (\citealt{Hollenbach1979}). Thus, observations of outflowing dust and molecules may provide an interesting diagnostic to break the degeneracies among the various models purporting to explain the presence of high velocity cold gas embedded in hot winds.

Moreover, the transport of dust in galactic winds may explain observations of dust in the CGM. SDSS extinction maps reveal the ubiquity of dust in CGM for galaxies above 10$^{10}$\,\Msun\ \citep{Zahid2013}, from distances of 20\,kpc to Mpc \citep{Menard2010}. The detection of dust at high redshift; i.e., by ALMA only 200\,Myr after the Big Bang (redshift 8.4; \citealt{Laporte2017}) and by JCMT in the Hubble Deep Field (redshift $\sim$4; \citealt{Hughes1998}), far before the evolution of low mass stars to the AGB phase, suggests supernova remnants act as major dust factories. Indeed, the galactic supernova remnant Cas A plausibly has produced of order 1\,\Msun\ of dust (\citealt{De2017}; \citealt{Bevan2017}; \citealt{Priestley2019}).

In this work, we extend earlier studies of ``warm'' (10$^4$\,K) gas entrainment by exploring the evolution of ``cold'' (10$^3$\,K) clouds in a ``hot'' (10$^6$\,K) wind; moreover we extend dust survival studies from high Mach cases appropriate for the interstellar medium to the transonic case relevant to the launching of galactic winds into the CGM. In \S \ref{sec:Methods}, we describe the numerical methods employed. In \S \ref{sec:Results}, we present our results. In \S \ref{sec:Discussion}, we discuss the relation of our results to previous work and highlight where future progress may be made. We summarize and conclude in \S \ref{sec:Conclusions}.


\section{Methods}
\label{sec:Methods}
\subsection{Numerical Setup}
\label{subsec:NumericalSetup}
We performed our simulations with the Eulerian grid code FLASH 4.2.2 (\citealt{Fryxell2000}; \citealt{Dubey2008}) using the unsplit staggered mesh solver (\citealt{Lee2009}; \citealt{Lee2013}) to solve the compressible, inviscid fluid equations, including radiative cooling as a sink term. We model dust as passive ``tracer'' particles to gain insight into the evolutionary history of clouds subjected to a hot wind (cf. \S\ref{sec:DustModeling} for details). To reduce computational expense, we implemented a ``cloud-tracking'' subroutine in FLASH (\S\ref{sec:CloudTracking}). See Table \ref{tab:flashParams} for a list of the particular numerical settings used in our simulations.
\begin{table}
\caption{Simulation parameters.}
\label{tab:flashParams}
\begin{tabular}{ l c }
\hline
 Parameter Name & Parameter Value \\ 
 \hline
 RiemannSolver & HLLC \\  
 slopeLimiter & mc \\
 order & 2$^a$ \\
 charLimiting & False$^b$ \\
 use\_hybridOrder & True \\
 hybridorderkappa & 0.5 \\
 cfl & 0.2 \\
\hline
\end{tabular}

$a$. MUSCL-Hancock,\\
$b$. Reconstruction in the primitive variables.
\end{table}

\begin{figure}
  \begin{center}
    \leavevmode
    \includegraphics[width=0.95 \linewidth]{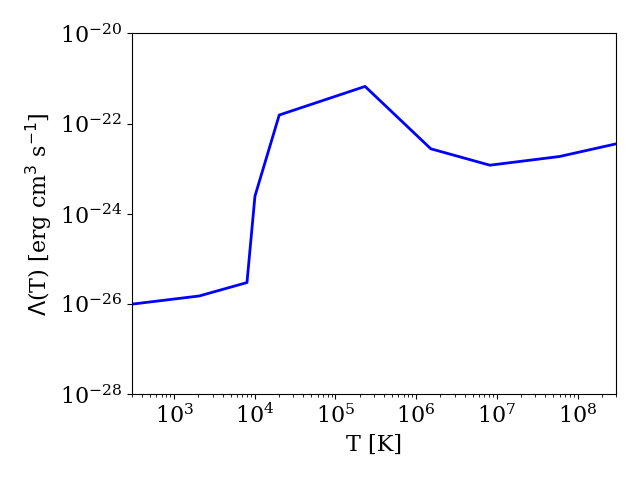}
\caption[]{Cooling curve used in the presented simulations, utilizing a piecewise power law fit to \citet{Sutherland1993} above \tenFour\,K and \citet{Dalgarno1972} below.}
\label{fig:CoolCurve}
\end{center}
\end{figure}

To integrate optically thin radiative cooling into our simulations we utilize the \citet{townsend2009} exact integration scheme\footnote{The Townsend scheme is ``exact'' because the cooling equation $\p{e}{t} = -n_H^2 \Lambda$ can be integrated analytically if $\Lambda(T)$ is a power law (or other integrable dependence).}\footnote{Note that the Townsend scheme has no timestep restriction and avoids overcooling inaccuracies of implicit solvers.} (implemented in FLASH, cf. test in \citealt{Farber2018}), with a piecewise power law fit to the \citet{Sutherland1993} cooling curve down to $10^4\,$K, which we extend down to 300$\,$K using the cooling curve of \citet{Dalgarno1972}. That is, we compute the cooling rate (shown in Figure \ref{fig:CoolCurve}) as 

\begin{table}
\caption{\label{tab:cooling}Piece-wise Power Law Fit to the Cooling Curve.}
\begin{tabular}{ c c r }
\hline
 Temperature Range & Coefficient (erg cm$^3$ s$^{-1}$) & Index \\ 
 \hline
 300\,K $\leq$ T $<$ 2000\,K & $10^{-26}$ & 0.2 \\  
 2000\,K $\leq$ T $<$ 8000\,K & $1.5 \times 10^{-26}$ & 0.5 \\
 8000\,K $\leq$ T $<$ $10^4$\,K & $3 \times 10^{-26}$ & 19.6 \\
 $10^{4}$\,K $\leq$ T $< 2 \times 10^{4}$\,K & $2.4 \times 10^{-24}$ & 6 \\
 $2 \times 10^{4}$\,K $\leq$ T $< 2 \times 10^{5}$\,K & $1.5438 \times 10^{-24}$ & 0.6 \\
 $2 \times 10^{5}$\,K $\leq$ T $< 1.5 \times 10^{6}$\,K & $6.6831 \times 10^{-22}$ & -1.7 \\ 
 $1.5 \times 10^{6}$\,K $\leq$ T $< 8 \times 10^{6}$\,K & $2.7735 \times 10^{-23}$ & -0.5 \\
 $8 \times 10^{6}$\,K $\leq$ T  $< 5.8 \times 10^{7}$\,K & $1.1952 \times 10^{-23}$ & 0.22 \\
 $5.8 \times 10^{7}$\,K $\leq$ T & $1.8421 \times 10^{-23}$ & 0.4 \\ 
 
\hline
\end{tabular}
\end{table}
\begin{equation}
\Lambda(T) = c_{k} \left(\frac{T}{T_k} \right)^{\alpha_k}
\end{equation}
where $\Lambda(T)$ is in the units of erg cm$^{3}$ s$^{-1}$, $k$ indicates the temperature range, and $\alpha_{k} = $log$\left(\frac{\Lambda_{k+1}}{\Lambda_k}\right)$ / log$\left(\frac{T_{k+1}}{T_{k}}\right)$; cf. Table \ref{tab:cooling} for the values of the coefficients, indices, and temperature ranges we employed. The above cooling curve assumes gas of solar metallicity. We enforce a temperature floor of \Tcl\ and we turn off cooling above 0.6 \Twind\footnote{Note that our temperature floor crudely approximates the effects of heating terms (e.g., cosmic rays, photoionization from the UV metagalactic background and young stars, and photoelectric heating).}, \Twind = $\chi$ \Tcl is the temperature of the hot wind.

\subsection{Initial Conditions}
We initialize a stationary, spherical cloud of radius \rcl, temperature \Tcl, and density \ncl\ in a hot homogeneous wind with a density contrast $\chi$ flowing at a $\sim$transonic Mach number, \Mach\ = 1.5. For the specific physical and numerical values used in our suite of simulations, see Table \ref{tab:ListOfRuns}.\footnote{Note that we utilized \ncl\ = 0.1 \cc\ in all runs. However, the only dynamically important quantities are, e.g., $\chi$, $\mathcal{M}$, and $c_{\rm s} t_{\rm cool}/r_{\rm cl}$ \citep{Scannapieco2015}, so one can renormalize, for instance, the number densities accordingly.}
Unless otherwise noted, we shall discuss the $\chi = 10^3$, \Tcl\ = 10$^3\,$K, \rcl/\dcell\ = 16, \rcl\ $\in$ (0.01, 10, 100)\,pc ``fiducial'' simulations (see Appendix~\ref{App:ResolutionConvergence} for convergence tests). 
\providecommand{\fourEfour}{$4 \times 10^{4}$}

\begin{table*}
\caption{\label{tab:ListOfRuns}Run parameters.}
\begin{tabular}{rrrrrrrrrrr}
\toprule
$\chi$ & T$_{\rm cl}$\,(K) &  $r_{\rm cl}$ (pc) &  $t_{\rm cc}$ (Myr) &  $t_{\rm cool,max}$ (Myr) &  $t_{\rm cool,minmix}$ (Myr) & $t_{\rm cool,mix}$ (Myr) &  $t_{\rm cool,wind}$ (Gyr) &  $r_{\rm cl}$/$d_{\rm cell}$ & L ($r_{\rm cl}$) & status \\
\midrule
                     10$^3$ & 400 &                  3 &                0.83 &                  210 &                     2.3 &                  2.6 &                0.097 &                            8 &      256 &     destroyed \\
                     10$^3$ & 400 &                 30 &                 8.3 &                  210 &                     2.3 &                  2.6 &                0.097 &                            8 &      256 &  warm survived \\
                     10$^3$ & 400 &                100 &                  28 &                  210 &                     2.3 &                  2.6 &                0.097 &                            8 &      256 &  warm survived \\
                    10$^3$ & 10$^3$ &     0.1 &               0.018 &                   69 &                     1.8 &                 0.32 &                  1.1 &                            8 &      256 &     destroyed \\
                    10$^3$ & 10$^3$ &                0.1 &               0.018 &                   69 &                     1.8 &                 0.32 &                  1.1 &                           16 &       64 &     destroyed \\
                    10$^3$ & 10$^3$ &                  2 &                0.35 &                   69 &                     1.8 &                 0.32 &                  1.1 &                            8 &      256 &     destroyed \\
                    10$^3$ & 10$^3$ &                 10 &                 1.8 &                   69 &                     1.8 &                 0.32 &                  1.1 &                            8 &      256 &     destroyed \\
                    10$^3$ & 10$^3$ &                 25 &                 4.4 &                   69 &                     1.8 &                 0.32 &                  1.1 &                            8 &      256 &     destroyed \\
                    10$^3$ & 10$^3$ &                 50 &                 8.7 &                   69 &                     1.8 &                 0.32 &                  1.1 &                            8 &      256 &  warm survived \\
                    10$^3$ & 10$^3$ &                100 &                  18 &                   69 &                     1.8 &                 0.32 &                  1.1 &                            8 &      256 &  warm survived \\
                    10$^3$ & 10$^3$ &                100 &                  18 &                   69 &                     1.8 &                 0.32 &                  1.1 &                           16 &       64 &  warm survived \\
                    10$^3$ & 10$^3$ &                500 &                  87 &                   69 &                     1.8 &                 0.32 &                  1.1 &                            8 &      256 &      survived \\
                    10$^3$ & 10$^3$ &               1000 &                 180 &                   69 &                     1.8 &                 0.32 &                  1.1 &                            8 &      256 &      survived \\
                    10$^3$ & 10$^3$ &               1000 &                 180 &                   69 &                     1.8 &                 0.32 &                  1.1 &                           16 &       64 &      survived \\
                   10$^3$ & 10$^4$ &                3.8 &                0.21 &                  1.1 &                     2.5 &                  1.6 &                   51 &                            8 &      256 &     destroyed \\
                   10$^3$ & 10$^4$ &                380 &                  21 &                  1.1 &                     2.5 &                  1.6 &                   51 &                            8 &      256 &      survived \\
                   10$^3$ & $4 \times 10^4$ &                320 &                 8.7 &                   47 &                      47 &                   69 &                  150 &                            8 &      256 &     destroyed \\
                   10$^3$ & $4 \times 10^4$ &              32000 &                 870 &                   47 &                      47 &                   69 &                  150 &                            8 &      256 &      survived \\
\bottomrule
\end{tabular}

Note that we used n$_{\rm cl}$ = 0.1 \cc\ in all simulations, whereas one may safely assume molecular clouds have higher densities (and hence (linearly)) smaller radii.
\end{table*}

\begin{figure}
  \begin{center}
    \leavevmode
    \includegraphics[width=0.95 \linewidth]{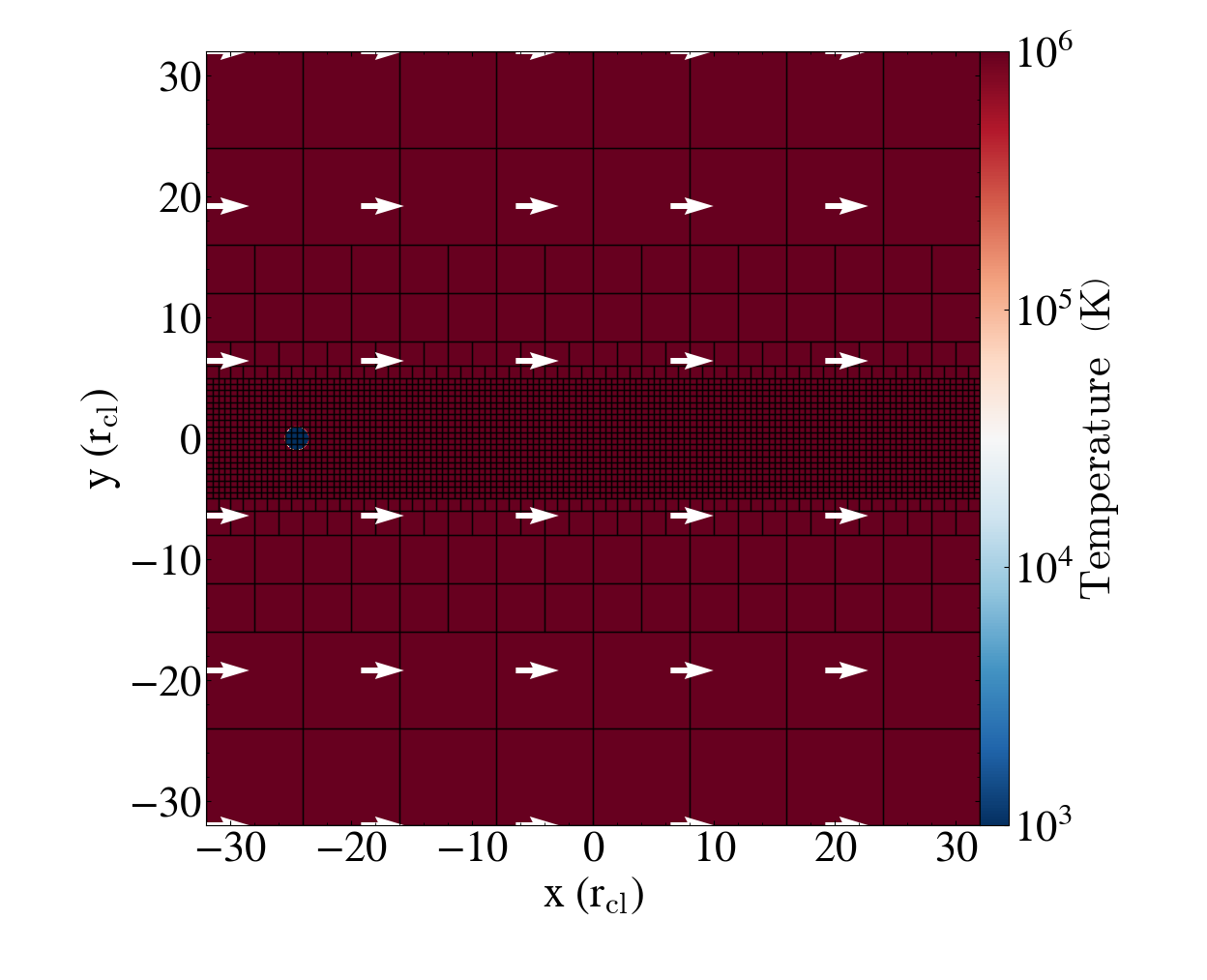}
\caption[]{Initial conditions showing the grid for one of the fiducial simulations. The pale blue dot indicates our cloud (see $x$=-26 \rcl, $y$=0). The white quivers indicate the velocity of the hot wind, which flows in from the -x boundary.}
\label{fig:GridIC}
\end{center}
\end{figure}

Our simulation domain consists of a cubical box with highest resolution for the interior $\sim$10 \rcl\ perpendicular to the wind, and uniform refinement in the direction of the wind. We apply outflow (zero-gradient) boundary conditions except for the $-x$ boundary where an inflow (wind) boundary condition is applied.\footnote{The inflow boundary condition is: $n_{\rm in} = n_{\rm w}, T_{\rm in} = T_{\rm w}, v_{\rm in} = v_{\rm w} = \mathcal{M} c_{s,w}$.} After $\pm$5 \rcl\ from the center of the cloud in directions perpendicular to the wind, we permit resolution to degrade as quickly as possible (note we use 8 cells per block, the FLASH default). See Figure \ref{fig:GridIC}. Thus, we simulate a large domain perpendicular to the wind direction at minimal computational expense, capturing the proper geometry of the bow shock and minimizing influence of the boundary conditions on the numerical solution (cf. \citealt{Scannapieco2015}; \citealt{martizzi2016}).

We position the initial center of the cloud to be (7.5, 0, 0) \rcl\ from the center of the inflow boundary. Moreover, we guarantee the front of the bow shock remains in the simulation volume (and that cloud material does not interact with the inflow boundary) by applying a buffer region of extent 4 \rcl\ from the inflow boundary as part of the cloud tracking scheme.

\subsection{Cloud-Tracking}
\label{sec:CloudTracking}
To reduce the amount of cloud material outflowing from the downstream face of our domain, we implemented a cloud tracking scheme in FLASH (\citealt{McCourt2015}; \citealt{Gronke2018}). To do so, we follow the evolution of a Lagrangian tracer concentration ($C$; \citealt{Xu1995}), which we initialize to $C = 1$ inside the cloud and zero elsewhere. This concentration variable is updated with the same numerical scheme as the other fluid variables, solving the conservation law

\begin{equation}
\p{\rho C}{t} + \bnabla \cdot (\rho C \textbf{u}) = 0
\end{equation}
which we use to compute the average cloud velocity

\begin{equation}
\langle u_{\rm cl}\rangle = \frac{\int_{x_{\rm min}}^{x_{\rm cl,0}} u_{x} C_{\rm cl} \rho\, \mathrm{d}V}{\int_{x_{\rm min}}^{x_{\rm cl,0}} C_{\rm cl} \rho\, \mathrm{d}V}
\end{equation}
where $x$ refers to the dimension in the direction of the initialized flow, $x_{\rm cl,0}$ is the $x$-coordinate of the initial center of the cloud, $x_{\rm min}$ is the $x$-coordinate of the minimum of the domain (bordering the inflow face), $u_{x}$ is the velocity along the x dimension, $C_{\rm cl} = C >$ max($C$)/3, and d$V$ is the cell volume.\footnote{Note the maximum of $C$ is taken every time step.}

Note that shifting to the cloud reference frame\footnote{subtracting $u_{\rm cl}$ from the inflow, grid, and particle velocities at every timestep.} (cf. \citealt{dutta2019}) not only reduces box size constraints to track cloud material but additionally reduces advection errors \citep{robertson2010}.

In our cloud tracking scheme, we added a feature to guarantee no cloud material hits the inflow boundary: if any material with $C > 0.1$ is flowing upstream within a buffer region of size 4 \rcl\ from the inflow boundary, we switch to a reference frame twice the speed of the fastest detected upstream flowing gas. Reference shifting is then paused until all detectable material exits the buffer region (downstream). Subsequently, we resume reference switching to the cloud reference frame as described above.

\subsection{Dust Model}
\label{sec:DustModeling}
We model dust as $10^6$ velocity tracer particles (cf. \citealt{Genel2013} for a discussion of velocity tracers compared to Lagrangian Monte Carlo tracers), which track the density and the temperature of the gas at each step, interpolated to the position of the particle. Initially, dust is randomly distributed within a cubical subvolume\footnote{See Appendix \ref{sec:DustInitialConditions} which shows the initial placement of the particles in this manner doesn't impact survival fraction more than $\sim$10\%.} of 0.75 \rcl\ (to ensure no particles are initialized outside the cloud due to discretization of the sphere) in an ``alive'' state. We set dust particles to a ``dead'' state if T $\geq$ \Tdest, which is a parameter we vary. In post-processing we determine the cumulative time each particle inhabits various bins of \Tdest to include the effect of dust surviving in hot gas for a finite `sputtering' time\footnote{Physically, the grains are eroded or `sputtered' during collisions with the thermal gas. Thus large grains diminish in size until they return to the gas phase.}.


\section{Results}
\label{sec:Results}

\begin{table}
\caption{\label{tab:definitions}Notation.}
\begin{tabular}{ l r }
\hline
 Name & Property \\ 
 \hline
 \lrg & 100\,pc \\  
 \med & 10\,pc \\
 \sml & 0.01\,pc \\
 cold & $\sim$10$^{3}$ K \\
 warm & $\sim$10$^{4}$ K \\
 hot  & $\sim$10$^{6}$ K \\
 \Tmix & \big(\Tcl$\,$\Twind\big)$^{1/2}$ \\
 \tmixCool & $t_{\rm cool}$ \big(T = \Tmix\big) \\
 \tmax & max\big($t_{\rm cool}$ (T < \Tmix)\big) \\
 \tminmix & $t_{\rm cool}$ \big($\sqrt{T(\rm min(t_{\rm cool})) \Twind}$\big) \\ 
 
\hline
\end{tabular}
\end{table}

\subsection{Three Evolutionary Paths for Cold Clouds}
\label{subsec:ThreePaths}
\begin{figure*}
  \begin{center}
    \leavevmode
    \includegraphics[width=\textwidth]{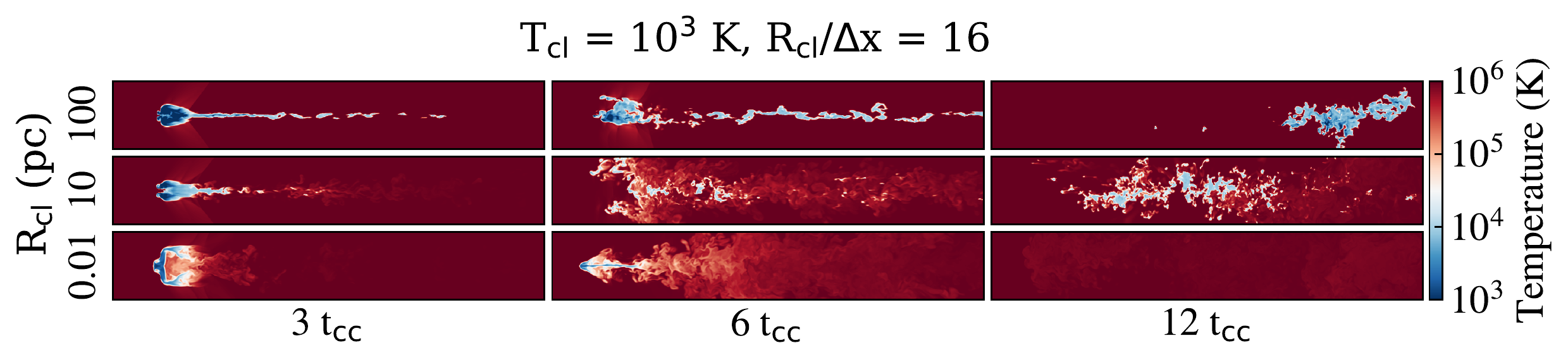}
\caption[]{Temperature slices along the z-axis for three different cloud sizes (marked on the left). Left-right shows time evolution. Top (100$\,$pc): survival of cold gas, middle (10$\,$pc): destruction of cold gas yet survival of warm gas, bottom (0.01$\,$pc): destruction. Note the simulation domain is much larger than the region shown; however, we zoom in on the cloud material (the highest resolution region). See Figure \ref{fig:SlicePlotZoom} for a zoom-in on the cloud material of interest.}
\label{fig:SlicePlotFull}
\end{center}
\end{figure*}

We begin the discussion of our results with the discovery that three evolutionary paths exist for cold (\Tcl\ $\sim$\tenThree\,K) clouds. In Figure \ref{fig:SlicePlotFull} we graphically illustrate the three characteristic evolutionary fates of cold clouds: (i) survival of cold gas, (ii) destruction of cold gas yet survival of warm gas, and (iii) complete destruction of cold and warm cloud material, via incorporation into the hot wind. 

Specifically, we show temperature slices (red-blue indicate hot-cold gas) in Figure \ref{fig:SlicePlotFull}, including simulations with $T_{\rm cl}=10^{3}\,$K and $r_{\rm cl}=100,\,10,\,0.01\,$pc (from top to bottom; moving forward we refer to these clouds as \lrg, \med, and \sml\ as defined in Table \ref{tab:definitions}). One can see that while the smallest cloud is being destroyed on a time-scale of $\sim$10\,$t_{\rm cc}$, the largest cloud clearly survives. The $r_{\rm cl}=10\,$pc cloud is in an interesting intermediate regime: at $t\gtrsim 6 t_{\rm cc}$ no cold ($T\sim$\tenThree\,K) gas remains yet the warm ($T\sim 10^4$\,K) phase persists. In contrast, previous work focused on \Tcl\ $\sim$\tenFour$\,$K clouds, finding the binary result of survival or destruction depending on if the cloud crushing time exceeded the cooling time of mixed gas. We have developed a new survival condition related to the three evolutionary paths (see \S \ref{subsec:SurvivalCriteria}), as well as a condition which sets when clouds transition from a destruction regime to a growth regime (for clouds that survive; see \S \ref{subsec:growthCondition}).

\begin{figure*}
  \begin{center}
    \leavevmode
    \includegraphics[width=\textwidth]{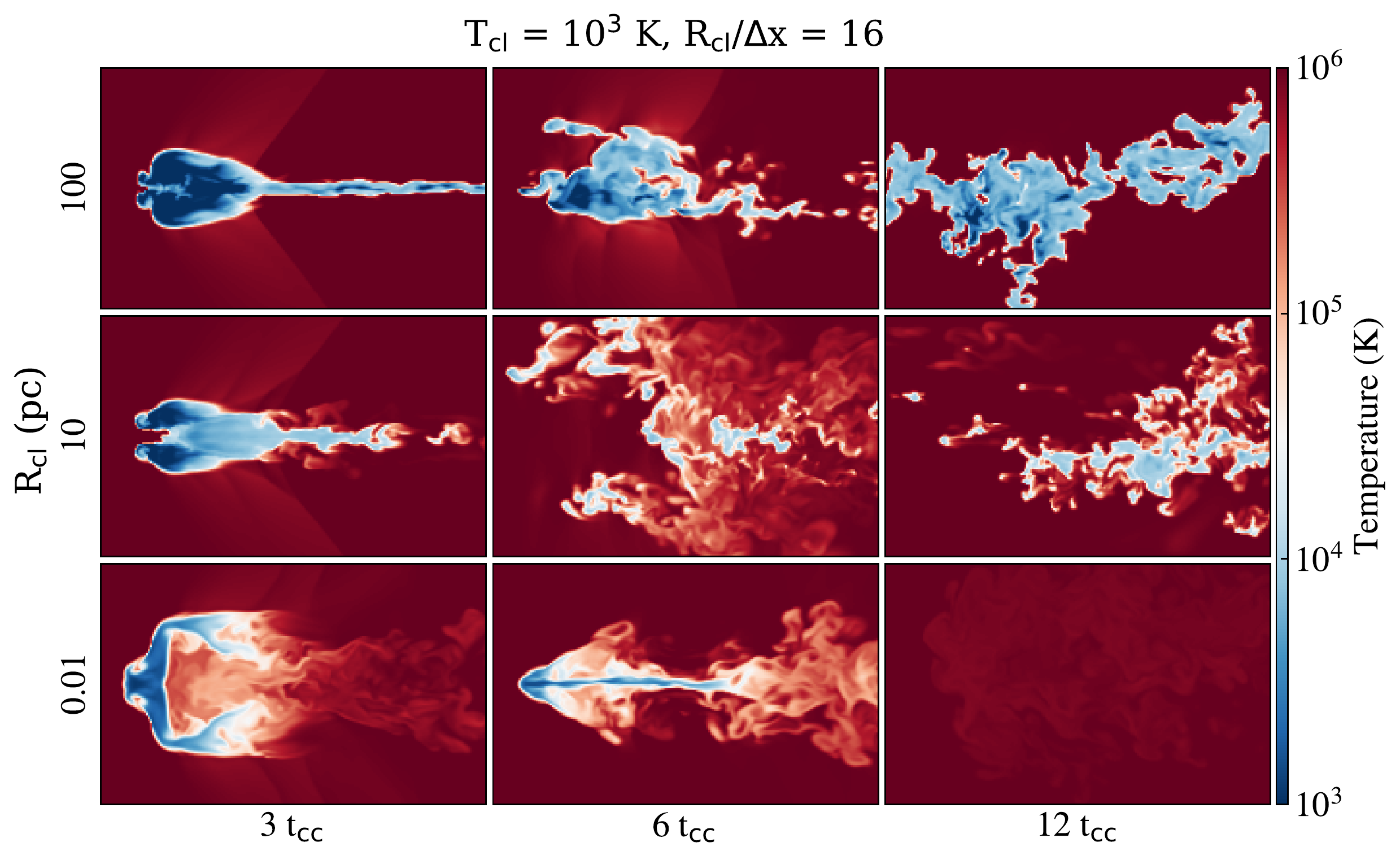}
\caption[]{Same as Fig. \ref{fig:SlicePlotFull} but zoomed in on cloud material.}
\label{fig:SlicePlotZoom}
\end{center}
\end{figure*}

Taking a closer look via Figure \ref{fig:SlicePlotZoom} we see hints of these three fates early in the evolution. At 3 \tcc\ \lrg\ has a more extended tail with cold gas (dark blue). At the same time \med\ has a shorter, wider tail and more warm gas at the head of the cloud (light blue). For \sml, in which case cooling is very inefficient, most of the tail has already been incorporated into the hot phase and the interior contains a growing fraction of hot gas.

At 6 \tcc\ (central column in Figure \ref{fig:SlicePlotZoom}) \lrg\ is nearly entrained ($\Delta v \lesssim 0.1 v_{\rm wind}$) and its cold gas is largely cocooned by warm gas. At the same time, \med\ only has cold gas remaining, which is vigorously mixing with the hot phase. \sml\ is not entrained yet at 6 $t_{\rm cc}$ (recall that the acceleration time is $t_{\rm drag}\sim \chi^{1/2}t_{\rm cc}\sim $32 $t_{\rm cc}$ for these simulations); we will analyze the entrainment process further in \S \ref{subsec:Sandman}.

At 12 \tcc\ the large cloud has collected its tail back into a semi-monolithic structure. The cold gas is now growing in mass as the destructive mixing rate has decayed with the dissipation of turbulence to the point radiative cooling dominates (see \S \ref{subsec:growthCondition} for the condition which sets the destruction vs. growth regime). \med\ has accumulated much warm gas but it is still in the process of merging (and conversely being incorporated into the hot phase for the smaller cloudlets). \sml\ has essentially been completely incorporated into the hot phase by this point.

\begin{figure*}
  \begin{center}
    \leavevmode
    \includegraphics[width=0.8 \textwidth]{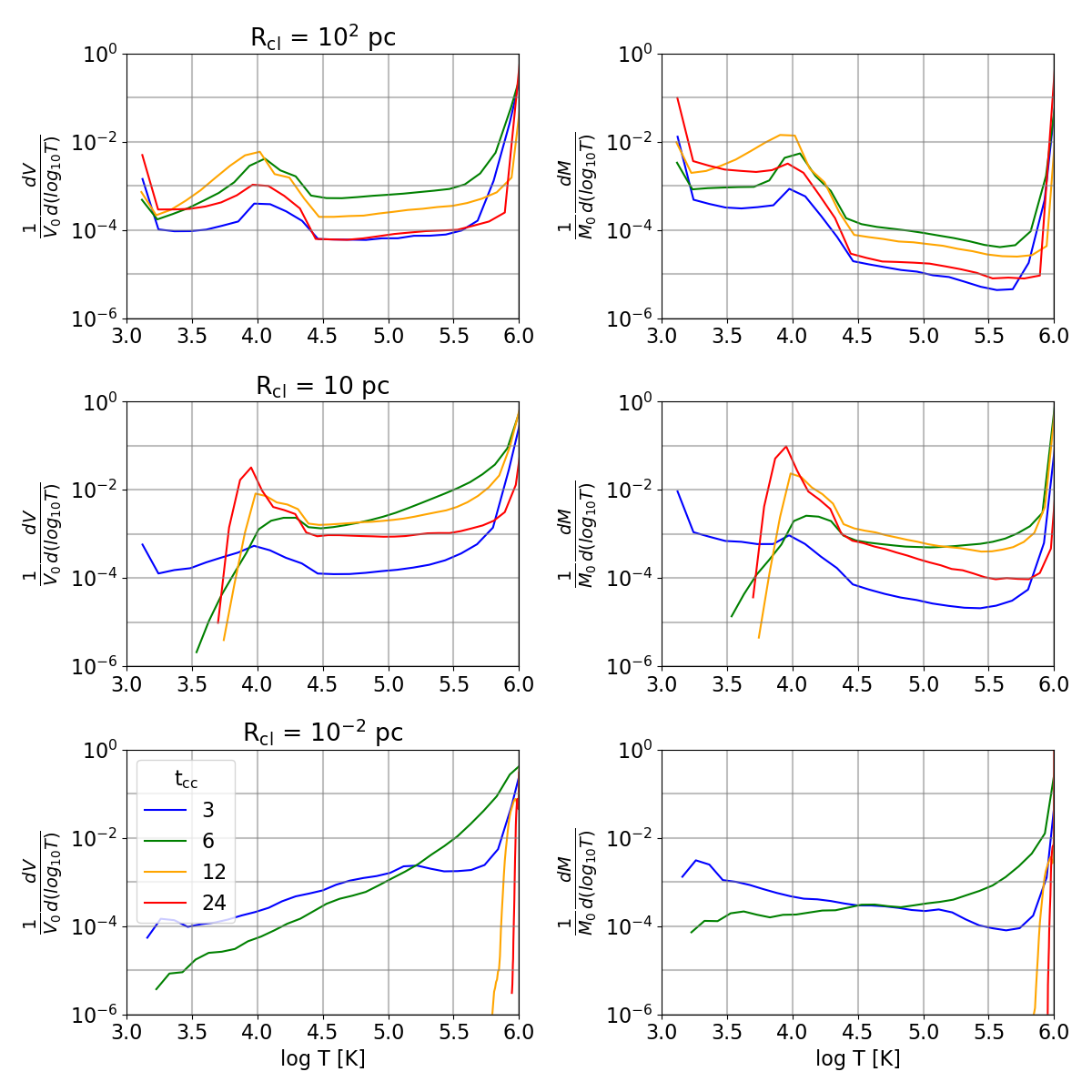}
\caption[]{Temperature distributions (left: volume, right: mass weighted). Top-bottom: \lrg\ (100\,pc), \med\ (10\,pc), and \sml\ (0.01\,pc) \Tcl\ = \tenThree$\,$K clouds. \med\ and \lrg\ clearly maintain a substantial amount of gas at \tenFour$\,$K.}
\label{fig:TemperatureHistogram}
\end{center}
\end{figure*}

The substantial growth of the warm phase hinted by the cocooning phenomenon is seen more clearly in the temperature probability distribution functions for volume (left) and mass (right) of Figure \ref{fig:TemperatureHistogram}. The blue, green, orange, and red curves respectively refer to the state at 3, 6, 12, and 24 \tcc. After a few \tcc, mixing fills out the intermediate temperature space between the cold and hot phases. While \sml\ rapidly disintegrates into solely hot wind material, \med\ and \lrg\ clearly demonstrate the growth of the \tenFour$\,$K (warm) phase as a peak distinct from the \tenThree$\,$K (cold) gas.

\begin{figure}
  \begin{center}
    \leavevmode
    \includegraphics[width=0.40\textwidth]{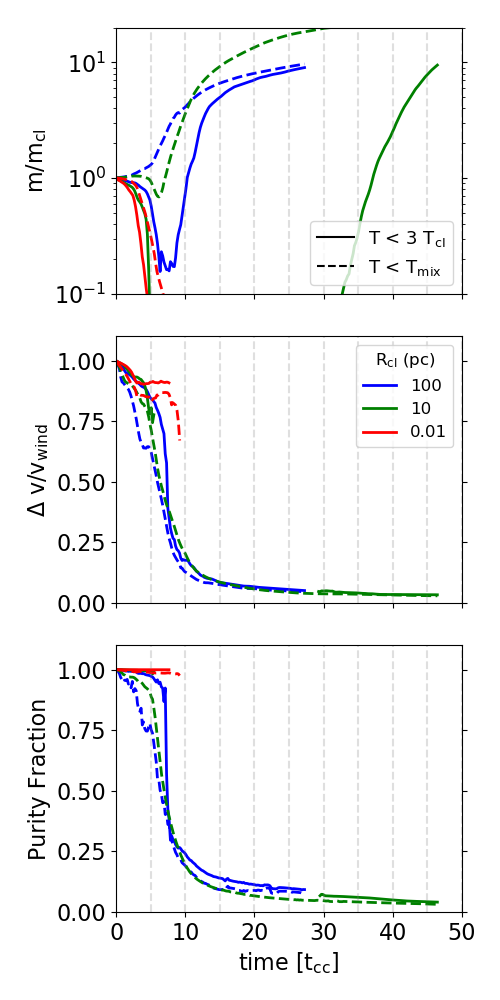}
\caption[]{Time evolution from top-bottom of: cloud mass, velocity shear between the cloud and the wind, and the fraction of original cloud material (median value of the tracer concentration). Solid lines use a cloud temperature maximum of 3 \Tcl\ (cold gas), while dashed lines use a cutoff at \Tmix\ (cold and warm gas). Blue, green, and red curves represent \lrg\ (100\,pc), \med\ (10\,pc), and \sml\ (0.01\,pc) clouds respectively. Note we smooth data every 0.2 \tcc\ (full data contains outputs every 0.02 \tcc) to improve presentation clarity.}
\label{fig:TimeSeries}
\end{center}
\end{figure}

Let us next take a more granular consideration of the full evolutionary history of the three characteristic clouds, as presented in Figure \ref{fig:TimeSeries}. From top to bottom we show time series of the cloud mass, velocity difference between the cloud and wind (``shear''), and fraction of gas which originated in the cloud (``purity'' fraction). That is, the purity fraction is the median value of the concentration $C$. We indicate \lrg\ (100\,pc) as blue curves, \med\ (10\,pc) as green curves, and \sml\ (0.01 pc) as red curves. In all the panels of Figure \ref{fig:TimeSeries}, solid curves indicate the cold phase (T < 3 \Tcl), and dashed curves indicate the combination of warm and cold phases (T < \Tmix). 

The top panel of Figure \ref{fig:TimeSeries} shows the mass evolution of the simulations shown in Figures \ref{fig:SlicePlotFull} \& \ref{fig:SlicePlotZoom} using two different temperature thresholds (note, we test convergence of the mass evolution in \S \ref{App:ResolutionConvergence}). We see that indeed \med\ loses its cold gas rapidly compared to \lrg\ (the rebirth of the cold gas at $\sim$30 \tcc\ is explained in \S \ref{subsec:growthCondition}). However, after an initial period of loss, \med\ attains more warm gas compared to \lrg\ as a result of rapid mass growth of the warm phase when the cloud is entrained, (cf. the $\Delta v$ panel, second row of Figure \ref{fig:TimeSeries}). The corresponding drop in purity fraction indicates the clouds are rapidly entrained due to efficient transfer of momentum from the hot wind to the cloud as a result of mixing and cooling (as opposed to small shards of cloud material being efficiently accelerated by ram pressure; cf. \citealp{Gronke2018,Abruzzo2021,Tonnesen2021}).

\Tcl\ = \tenThree\,K clouds clearly exhibit three evolutionary paths. However, what leads to this behavior which clearly differs compared to the well-studied \Tcl\ = \tenFour\,K case? We argue the presence of a local maximum in the cooling time at $\sim$\eightEThree$\,$K (as opposed to the $\sim$monotonic increase in cooling time with temperature above $\sim$\tenFour$\,$K) is responsible for a semi-stable warm phase. 

\subsection{Survival Criteria}
\label{subsec:SurvivalCriteria}
\subsubsection{Characteristic Cooling Time-scales}
\label{subsubsec:CoolingTimescales}

\begin{figure}
  \begin{center}
    \leavevmode
    \includegraphics[width=0.45\textwidth]{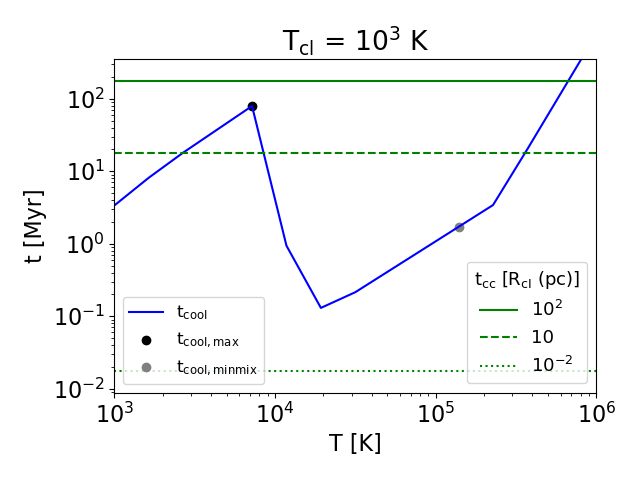}
\caption[]{Comparison of the isobaric cooling time (blue, solid) to the cloud crushing times (green) of \lrg\ (solid), \med\ (dashed), and \sml\ (dotted). We have indicated important cooling times with respect to the cloud crushing time which dictate the evolutionary result of the cloud. The maximum cooling time, \tmax, is indicated as a black point and must be shorter than the cloud crushing time for survival of cold cloud material. The minimum cooling time of the mixed gas, \tminmix, is marked as a grey dot and must be shorter than the cloud crushing time for any cloud material to survive.}
\label{fig:Timescales}
\end{center}
\end{figure}

In Figure \ref{fig:Timescales} we show the isobaric cooling time as a function of temperature assuming pressure equilibrium by the blue curve. The green curves indicate the cloud crushing time of the clouds with the solid, dashed, and dotted curves indicating \lrg, \med, and \sml\ respectively.  

The first time-scale to consider is the cooling time of the mixed gas as considered by \citet{Gronke2018} as survival criterion for $\sim 10^4\,$K gas
\begin{equation}
    t_{\rm cool,mix} \equiv t_{\rm cool} \big(T = \Tmix\big)
\label{eqn:tmixCool}
\end{equation}
with $T_{\rm mix}=\sqrt{T_{\rm wind}T_{\rm cl}}$ as derived in a simple model of turbulent mixing layers by \citet{Begelman1990}.

Second, we denote the longest cooling time of the gas between $T_{\rm mix}$ and $T_{\rm cl}$ as
\begin{equation}
    \tmax \equiv \mathrm{max}\big(t_{\rm cool} (T < \Tmix)\big).
\label{eqn:tmax}
\end{equation}
As such, \tmax\ = \tmixCool\ if the cooling time curve is monotonically decreasing, as is (roughly) the case for simulations truncating cooling at $10^4$\,K. 

And lastly, the cooling time of the gas with a temperature between the minimum cooling time and the hot gas:
\begin{equation}
    \tminmix \equiv t_{\rm cool} \big(\sqrt{T(\mathrm{min}(t_{\rm cool})) \Twind}\big).
\label{eqn:tminmix}
\end{equation}
Again, for a monotonic decreasing cooling curve, \tminmix\ = \tmixCool\ so we recover the traditional, binary survival criterion above $10^{4}$\,K. 
Note that in Eq.~\eqref{eqn:tminmix}, we also used the geometric mean as an average temperature for the mixed gas -- as was done for $t_{\rm cool,mix}$ above. While in practice the relevant cooling times of such mixed gas is some integrated quantity and, hence, dependent of the cooling curve between the two bordering temperatures \citep[see a discussion of the impact of shape of the cooling curve on $t_{\rm cool,mix}$ in][]{Abruzzo2021}, the geometric mean has proven to be a good approximation of it in the $\sim 10^4\,$K case \citep{Kanjilal2021}.
Note that $\tminmix = t_{\rm cool,mix}$ for $T_{\rm cl}\sim 10^4\,$K but the generalized form of Eq.~\ref{eqn:tminmix} allows us to consider lower $T_{\rm cl}$, and as we will see below, establish criteria for the different outcomes presented in \S \ref{subsec:ThreePaths}.



\subsubsection{Evaluation of Survival Criteria}
\label{subsec:SurvivalCriteriaEvaluation}

\begin{figure}
  \begin{center}
    \leavevmode
    \includegraphics[width=0.49\textwidth]{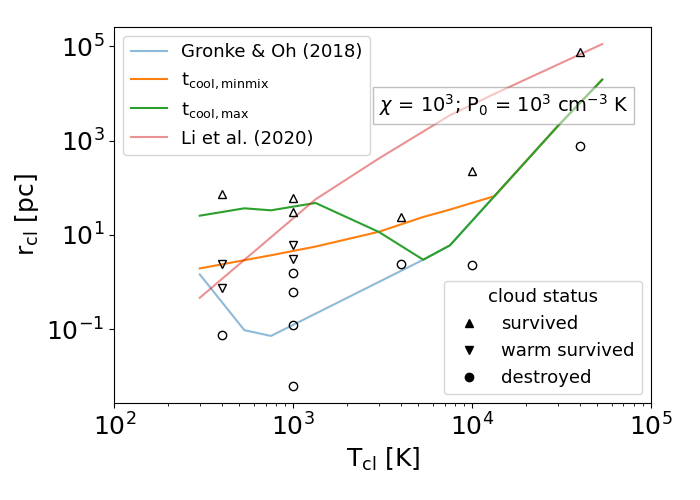}
\caption[]{Comparison of simulation results. Destruction of all cloud material is indicated as circles. Destruction of the cold phase and survival of the warm phase is indicated by upside down triangles. Survival of the cold phase is marked by triangles. The \tmax\ survival criterion (green) correctly predicts survival/destruction of cold gas while the \tminmix\ survival criterion (orange) predicts persistence of warm gas vs. complete destruction. We show the \tmixCool\ (\thot) survival criterion of \citet{Gronke2018} (\citealt{Li2020}) as blue (red) curves  for comparison.}
\label{fig:SurvivalAll}
\end{center}
\end{figure}

In the case of cooling down to 10$^{3}$\,K, three possible evolutionary results emerge. In the case that \tcc\ > \tmax, cold cloud material remains present throughout the simulation. When \tmax\ > \tcc\ > \tminmix, cold cloud material is heated as it gets mixed. The warm phase survives but is unable to cool efficiently down to the cloud temperature due to disruption via mixing (until the shear velocity driving such mixing sufficiently decays as the cloud is entrained, cf. \S \ref{subsec:growthCondition}). Note that when the cloud is entrained warm cloud material will eventually cool down to the cold phase (if dust survives, cf. \S \ref{subsec:DustSurvival}), so this distinction is regarding the evolutionary path rather than the final end state. Last, in the case that \tcc\ < \tminmix, all cloud material is incorporated into the hot wind, completely destroying the cloud (as in the classical adiabatic case). This picture is consistent with our three fiducial clouds as seen in Figure \ref{fig:SlicePlotZoom}. In the next section, we perform a suite of simulations to test the new criteria more rigorously.


In Figure \ref{fig:SurvivalAll} we indicate the fate of the clouds we simulated as a function of \Tcl\ and (initial) cloud radius (\rcl). The three evolutionary paths are cold phase survives (triangles), cold phase is destroyed but warm phase survives (upside down triangle), and complete destruction of cloud material (circles). Note that in comparison to Table \ref{tab:ListOfRuns}, we have renormalized the presented \rcl\ to show results at fixed pressure $P_0 / k_B$ = \tenThree\,\cc K. Recall all simulations presented are $\chi = 10^3$. We indicate the expected critical radii for cold phase survival (\tmax, green) warm phase survival (\tminmix, orange), as well as the \tmixCool\ criterion of \citet{Gronke2018,Gronke2020Cloudy} and the $\sim$\thot\ criterion of \citet{Li2020} for reference.

At \fourEfour$\,$K we recover the result of \citet{Gronke2020Cloudy}: survival for clouds with efficient mixing (\tmixCool/\tcc\ $\sim$\ 0.1) and destruction in the $\sim$adiabatic case \tmixCool/\tcc\ $\sim$\ 10. We find disagreement with the \thot\ criterion of \citet{Li2020} in the case of our \tenFour\,K cloud; however, this is unsurprising given the differences in the physics we include (note that with conduction, \citealt{Li2020} observe very little gas at intermediate temperatures since conduction removes temperature gradients whereas we neglect conduction; cf. \S \ref{sec:DiscussionComparisonCloudCrush} for a more detailed comparison with previous work).

Similarly, we observe disagreement with the \tmixCool\ criterion of \citet{Gronke2018,Gronke2020Cloudy} at \tenThree$\,$K. As the survival criteria are clearly separated at \tenThree$\,$K, we perform a higher sampling of cloud radii there to test the criteria. Allowing clouds a factor of two smaller/larger radii than the value indicated by the curve, we find agreement with \tmax\ and \tminmix. Interestingly, \citet{Li2020} is $\sim$consistent with \tmax\ at this temperature. We include a few representative cases at 400\,K for reference. Note that the hot phase in this case, \fourEfive\,K, should be short-lived and therefore the scenario is expected to be useful only for confirming the survival criteria theoretically as it is physically unlikely to be realized. Note the log-scaling so the smaller warm-survived case is only roughly a factor of two discrepant. We defer a study of the more realistic case of \Tcl\ = 400\,K with $\chi$ = \tenFour\ for future work.

\subsection{Condition for Mass Loss vs. Growth Phase}
\label{subsec:growthCondition}
Consider anew the cooling time curve in Figure \ref{fig:Timescales}. If \lrg\ has \tcc\ > \tmax, why does it take 10 \tcc\ before the \Tcl\ gas starts growing? 
A similar mass evolution, i.e., a mass loss phase followed by a growth phase has been observed in similar `cloud crushing' simulations before (e.g., \citealp{Gronke2018,Sparre2020,Kanjilal2021,Abruzzo2021}) and in fact might be related to different `survival criteria' (cf. \S \ref{sec:DiscussionComparisonCloudCrush}). 
Here, we provide a simple argument on the transition point between the mass loss and growth phase by comparing (once again) the mixing to the cooling time-scales (akin to what has been done in detailed studies focusing on the ISM context; \citealp{hennebelle1999,saury2014}).

Recall that since cooling becomes inefficient below \tenFour$\,$K, gas piles up at \tmax\ forming a cocoon of warm gas around the cold gas (cf. Figure \ref{fig:SlicePlotZoom}).
Hence, the mixing time experienced by the \tenThree$\,$K gas is then given by $\tmixCC \sim (\chi')^{1/2} L / u' \sim (T(\tmax) / T_{\rm cl})^{1/2} r_{\rm cl} / u'$ where we have assumed that the outer scale stays approximately constant, and $u'$ is the 3D turbulent velocity of the `cocoon', i.e., the $\sim T(\tmax)$ gas. More specifically, we computed the velocity dispersion of gas with T < \Tmix.

\begin{figure}
  \begin{center}
    \leavevmode
    \includegraphics[width=0.45\textwidth]{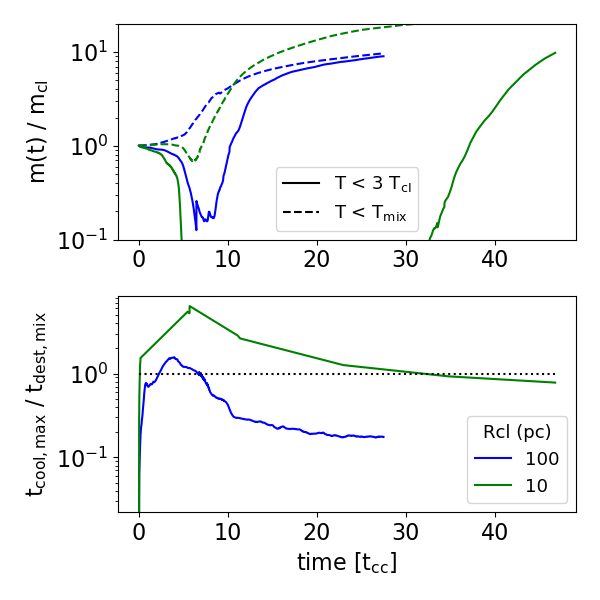}
\caption[]{Top panel: cold gas mass evolution. Bottom panel: evolution of the mixing time-scale $\tmixCC=(T(\tmax) / T_{\rm cl})^{1/2}r_{\rm cl}/u'$ where $u'$ is the 3D velocity dispersion of the $\sim 10^4\,$K `cocoon' (that is, T < \Tmix) compared to \tmax\ and $T(\tmax) / T_{\rm cl} = 8$ (cf. \S \ref{subsec:growthCondition} for details). In both cases, clouds grow in cold gas only after the mixing time-scale becomes long compared to the maximum cooling time.}
\label{fig:tmixCondition}
\end{center}
\end{figure}

Figure \ref{fig:tmixCondition} shows the evolution of the ratio of \tmixCC\ and the cooling time of the $\sim 10^4$\,K gas \tmax\ alongside the cold gas mass evolution (lower and upper panel, respectively). 
To evaluate \tmixCC\ we computed $u'$ as the 3D velocity dispersion of the $T<T_{\rm mix}$ gas.
If this mixing time is initially shorter than the cooling time, the cloud loses cold gas; however, when the \tmixCC\ becomes longer than cooling time the cloud enters the growth phase at $\sim$10 \tcc\ as observed for \lrg. Similarly, \med\ does not re-generate cold gas until $\sim$30 \tcc\ when the destructive mixing rate set by the turbulent velocity has dropped sufficiently low. 
Note that this means that for simulations where $\tmixCC > \tmax$ there is no `mass loss' phase and the cloud starts growing immediately. This is the case for some of our simulations: (i) \Tcl\ = 400\,K, $\chi$ = \tenThree, \rcl\ = 120\,pc; (ii) \Tcl\ = 400\,K, $\chi$ = \tenThree, \rcl\ = 40\,pc; (iii) \Tcl\ = \tenThree\,K, $\chi$ = 10$^2$, \rcl\ = 30\,pc.

Figure \ref{fig:tmixCondition} suggests a good agreement between the time at which our simulated clouds enter the growth phase with the \tmixCC/\tmax\ = 1.
This is somewhat surprising since we do only barely resolve the turbulence within the `cocoon' and, thus, the mixing of the $\sim 10^3\,$K gas seeds. However, as we do resolve the outer scale ($\sim r_{\rm cl}$), and the mixing rate is set by the motion on this scale, the (re)appearance of cold gas might be robust (cf. discussion in \S \ref{sec:DiscussionCaveats}). More detailed, high resolution simulations targeting this process are necessary to study this further. For instance, an interesting avenue for future work is determining the evolution of the turbulent velocity (cf. \citealp{Ji2018,Fielding2020,tan2021radiative} for the plane-parallel case) \textit{ab initio}, which would help determine the necessary simulation runtime to capture the mass growth phase.


\subsection{Comparison between warm and cold clouds}

In the following, we include simulations of warm clouds that survive \Tcl\ $\in$ (\tenFour$\,$K, \fourEfour$\,$K) to compare properties of our cold clouds with previous work.

\begin{figure}
  \begin{center}
    \leavevmode
    \includegraphics[width=0.34\textwidth]{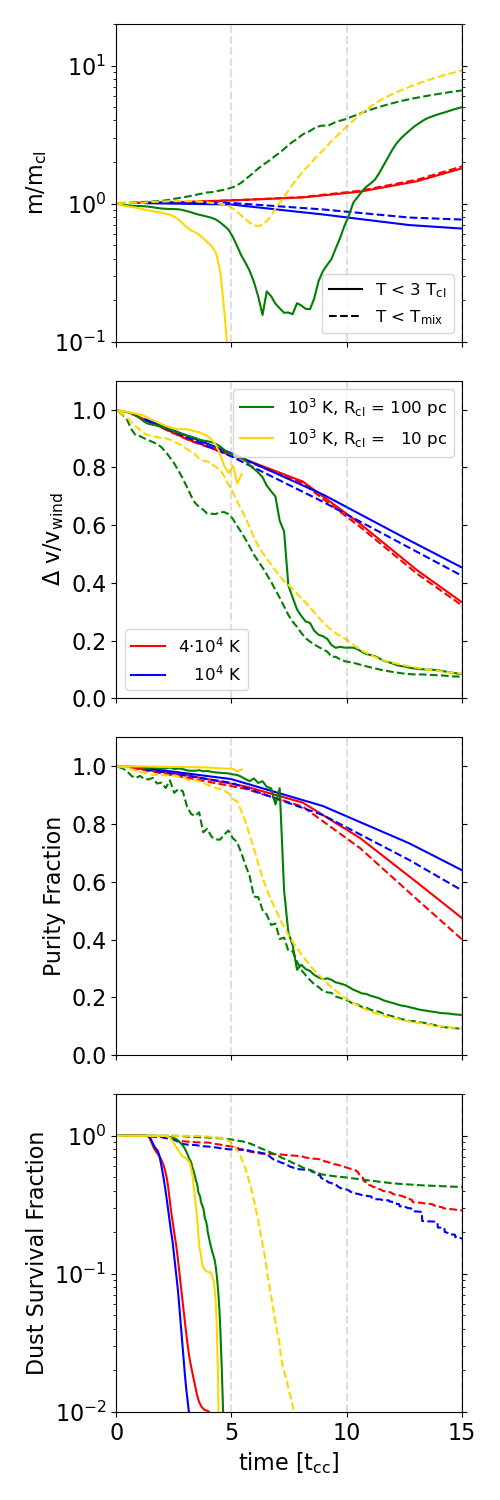}
\caption[]{Time evolution from top-bottom of: cloud mass, velocity shear between the cloud and the wind, the fraction of original material below the cloud temperature, and the fraction of dust particles that stayed below the cloud temperature. Solid lines use a cloud temperature maximum of 3 \Tcl, while dashed lines use a cutoff at \Tmix. Red, blue, green, and gold respectively refer to \Tcl = \fourEfour, \tenFour, \tenThree, and \tenThree\,K clouds with green and gold corresponding to \lrg\ and \med\ respectively. Note we smooth data every 0.2 \tcc\ (full data contains outputs every 0.02 \tcc) to improve clarity of presentation. The \tenFour and \fourEfour\,K clouds use \tmixCool/\tcc\ $\sim$ 0.08 as in \citep[][]{Gronke2018,Abruzzo2021} and \tminmix/\tcc\ $\sim$ 0.01 and 0.1 for the $R=100$\,pc and $10\,$pc cloud, respectively.}
\label{fig:TimeSeriesHotClouds}
\end{center}
\end{figure}

We now return to compare the time series for warm and cold clouds in Figure \ref{fig:TimeSeriesHotClouds}. In this case, the red, blue, green, and gold curves refers to the \Tcl\ = \fourEfour\,K, \tenFour\,K, \lrg, and \med\ clouds. From top-bottom we show the mass evolution, shear velocity, purity fraction, and the dust survival fraction (discussed in more detail below). Both of the \Tcl\ = $10^3\,$K clouds evolve much more rapidly toward their saturated state compared to warm clouds included for reference, a point we turn to in the next subsection. 

Recall that we model dust by initializing 10$^6$ velocity tracer particles inside the cloud. The bottom panel of Figure \ref{fig:TimeSeriesHotClouds} indicates the fraction of particles that remain in cloud material (the ``survival'' fraction); we subsequently consider finite sputtering time in \S \ref{subsec:DustSurvival}. Assuming a destruction temperature \Tdest\ = 3 \Tcl\ we find complete destruction of dust. However, permitting dust to survive up to \Tmix\ we find roughly 10s\% dust survives for \lrg, while still no dust survives for \med.

Interestingly, the dust survival fraction is largest for \lrg, likely due to the rapid growth of the warm phase (despite the destruction of all original cold gas, cf. \S\ref{subsec:Sandman}) and lowest (excluding \med\ which is rapidly mixed away) for the intermediate temperature \Tcl\ = $10^4\,$K cloud which takes longest to be entrained. We have shown only the first 15 \tcc\ of evolution to highlight the correlation of the evolution of the dust survival fraction with that of the shear velocity. Note, however, that the warm clouds have essentially complete dust destruction as they take roughly 30 \tcc\ to be entrained and dust is continuously destroyed as cloud material gets mixed with the hot wind. However, once a cloud is entrained, the dust survival fraction saturates and \lrg\ doesn't significantly change it's survival fraction (cf. Appendix \ref{App:DustSurvivalFractionConvergence}). In any case, we show the dust survival fraction with the full simulation time for the cold clouds in \S \ref{subsec:DustSurvival}.

\subsection{Rapid Entrainment of Cold Gas}
\label{subsec:Sandman}
\begin{figure}
  \begin{center}
    \leavevmode
    \includegraphics[width=0.4\textwidth]{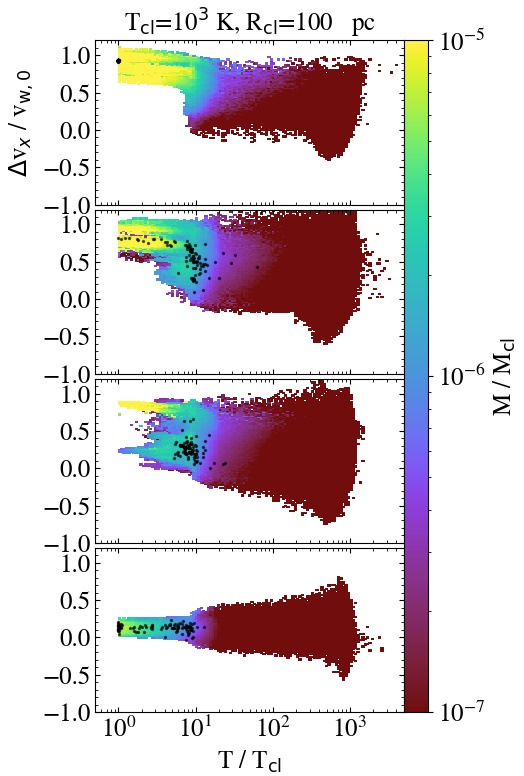}
\caption[]{Phase plots of the velocity shear (in the wind direction) between the molecular and hot phases vs. temperature. From top to bottom we show the state at 3.2, 6.5, 8.2, and 12.9 \tcc, with times selected for clarity of the particle evolution. Tracer particles are overplotted as black dots and reveal entrainment proceeds via heating up cloud material during the mixing process and the mixed entrained cloud material cools back down to the cold phase.}
\label{fig:Sandman}
\end{center}
\end{figure}

As we saw in Figure \ref{fig:TimeSeries}, the surviving clouds are accelerated on a time-scale of $\sim$ 10\tcc\ $\ll t_{\rm drag} \sim 30$ \tcc. The nature of this rapid entrainment is elucidated in Figure \ref{fig:Sandman}, which shows the phase space of shear between the wind and cloud as a function of temperature of the gas. Yellow (purple) colors indicate large (small) masses in a bin indicative of cloud (wind) material. Black dots indicate the phase location of tracer particles (note we only show one thousand for clarity), which indicate the original cloud material is heated into the warm phase and entrained as it mixes with the hot wind. That is, the top panel at 3.2 \tcc\ shows the particles still inside the original cloud. The second panel from top at 6.5 \tcc\ shows particles evolving towards zero shear as they are mixed, exchanging momentum with the hot wind. The second panel from bottom at 8.2 \tcc\ shows most particles are now entrained in the mixed gas. The bottom panel at 12.9 \tcc\ shows that once the cloud material is entrained, it cools back down to the cold phase.

That is, none of the original material remains in the cold phase (in agreement with \citealt{Tonnesen2021} at different scales, cf. their Figure 11); as original material is mixed, hot material that was mixed and entrained cools and replaces the original material. For \lrg, this process occurs sufficiently rapidly that there is always cold material, although after a relatively short period of time none of it belonged to the original cloud.

Although this suggests that instantaneous destruction of dust is efficient, modifying the sputtering time or destruction temperature can dramatically alter the survival fraction of dust, a point we turn to next.

Note that this process implies that for the rapid entrainment the `cocoon' of warm gas is necessary (which is not the case for the $T_{\rm cl}\sim 10^4\,$K clouds, cf. Figure \ref{fig:TimeSeries}). It forms quickly because of the short cooling time of warm material, and is subsequently easier to accelerate due to a lower over overdensity. Mixing and continuous cooling to lower temperatures -- as described above -- ensures that the cold gas is also successfully entrained.

\subsection{Dust Survival}
\label{subsec:DustSurvival}

\begin{figure}
  \begin{center}
    \leavevmode
    \includegraphics[width=0.45\textwidth]{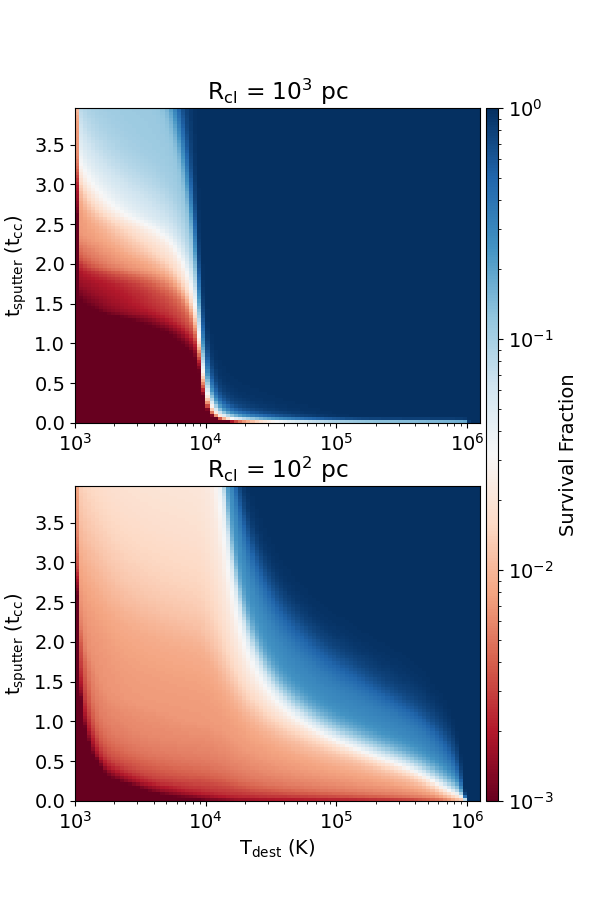}
\caption[]{Dust survival fraction dependence on destruction temperature ($x$-axis) and sputtering time ($y$-axis), measured at the final output for each run ($\sim$25 \tcc\ for \lrg\ \& 50 \tcc\ for \med). Blue (red) colors indicate near complete survival (destruction). Top (bottom) panel: \lrg\ (\med). Note the very strong cutoff at $\sim10^{4}$\,K for short sputtering times in the top panel.}
\label{fig:DSF_TdustdeathBracket}
\end{center}
\end{figure}

Since molecular H$_2$ requires re-formation on dust grains, the survival of dust plays a crucial role in the possible association of our coldest gas with a molecular phase. In this section, we consider dust destruction with a simple model that nevertheless brackets more involved modeling depending on the grain size distribution and grain species (see \S \ref{sec:DiscussionCaveats} for additional discussion).

Recall we define dust destruction conservatively such that dust is considered to be immediately destroyed when it encounters hot gas T $\geq$ \Tdest. To bracket the range of possible dust survival, we present in Figure \ref{fig:DSF_TdustdeathBracket} the dust survival fraction's dependence on \Tdest\ and sputtering time. Note that we measure the dust survival fraction using the full simulation time for each run $\sim$ 25 (50) \tcc\ for \lrg\ (\med). That is, one can relax our conservative definition of dust destruction by supposing dust can exist for some duration of time $\geq$ \Tdest\ before actually being destroyed (as in the case of thermal sputtering for example). We see that $10^{4}\,$K acts as a threshold for dust survival in the \lrg\ case (particularly for the case of instantaneous destruction, with zero sputtering time) and most of the dust survives if it can survive up to \Tdest\ = $10^{5}\,$K or several \tcc\ in the ``hot'' phase. In other words, large cloud dust survival depends primarily on \Tdest\ and relatively insensitive to the sputtering time, whereas smaller clouds' dust survival depends more crucially on the sputtering time.

Moreover, this suggests dust that mixes out of the cloud remains predominately in the mixed phase, which more moderate temperature and higher density might provide more protection from FUV induced evaporation than if dust immediately inhabited the hot wind phase.


\section{Discussion}
\label{sec:Discussion}
We begin the discussion of our results by comparing to previous work, specifically cloud crushing simulations of \Tcl\ = \tenFour\,K clouds (\S\ref{sec:DiscussionComparisonCloudCrush}) and  studies of the survival of cold gas \& dust in cloud crushing setups (\S~\ref{sec:DiscussionComparisonDust}). Thereafter, we discuss the implications of our work regarding observations of galactic winds, fountain flows, observations of dust and cold gas in CGM (\S \ref{sec:DiscussionGW}), and star formation in the jellyfish tails of ram pressure stripped galaxies (\S \ref{sec:DiscussionJelly}). We conclude this section by discussing potential caveats to our conclusions (\S \ref{sec:DiscussionCaveats}).

\subsection{Comparison to previous work}
\subsubsection{Cloud Crushing}
\label{sec:DiscussionComparisonCloudCrush}
Recent work on the cloud crushing problem has fixated on the survival criterion for \tenFour\,K clouds. While \citet{Li2020} \& \citet{Sparre2020} find their simulated clouds survive based on the cooling time of the hot wind, \citet{Abruzzo2021} \& \citet{Kanjilal2021} find general agreement with the \tmixCool\ condition of \citet{Gronke2018}. We attribute our warm clouds surviving with the \tmixCool\ condition as related to evolving these clouds in a sufficiently large box for a sufficiently long duration to capture the growth regime of our warm clouds. 

Regarding our disagreement with \citet{Sparre2020} who perform very similar simulations to ours, a number of departures could explain our different criteria depending on mixed vs. hot gas. As noted by \citet{Kanjilal2021}, all simulations resulting in a \tmixCool\ condition were performed with Eulerian grid codes whereas the \thot\ condition was found with Lagrangian-Eulerian codes. Eulerian codes produce extra mixing due to numerical diffusion. However, as \citet{Gronke2020Cloudy} found converged results for $\mathcal{M}\sim 1.5$ and numerical diffusion decreases with resolution, we expect our different discretization methods are not a significant concern. 

Another difference that may affect the results of \citet{Sparre2020} is their use of periodic boundary conditions\footnote{Note that the periodic boundary conditions transverse to the wind axis could mitigate the cooling flow onto the cloud and tail \citep[cf. figure 3 for the impact of the orthogonal boundary conditions][]{Gronke2018}.} and a relatively small numerical domain (64 \rcl\ in the flow direction and 16 \rcl\ transverse). However, we expect these numerical differences to play at most a minor role, as \citet{Sparre2020} state that their cloud material remains within their box during their simulated time.

A possibly more significant difference between our works is related to the temperature floor: we require the minimum gas temperature to be \Tcl\ in our simulations whereas \citet{Sparre2020} enforce \Tcl/2 (5000 K). If their clouds cool to \Tcl/2 rapidly this would significantly decrease the expected cloud survival radius (cf. Figure \ref{fig:SurvivalAll}). Moreover, the lower cooling floor may effectively increase their ``initial'' overdensity by a factor of two, implying a somewhat longer drag time and hence a longer phase in the destruction regime. Similarly, we employ a tabulated power-law cooling function whereas they use a rather more sophisticated method, decomposing cooling into a primordial component \citep[][]{katz1996damped}, Compton cooling off the cosmic microwave background \citep[][]{vogelsberger2013model}, and metal-line cooling using Cloudy models \citep[][]{ferland20132013} with a metagalactic UV background \citep[][]{faucher2009new}.

Yet the clearest and likely most significant departure of this work from \citet{Sparre2020} is the different definitions of cloud destruction we adopt; whereas we evolve our clouds for at least 25 \tcc\ independent of the cold mass content, \citet{Sparre2020} consider a cloud to be destroyed if its mass is not growing at 12.5 \tcc, and it is clear some of these ``destroyed'' clouds would survive under our criterion (cf. their 47\,pc cloud in Figure A2). That is, our simulations may produce the same results if interpreted uniformly.

As \citet{Li2020} include a number of additional physics we neglect, comparison is more difficult. Similar to \citet{Sparre2020}, possibly the largest departure between our works is related to our different definitions of cloud destruction; \citet{Li2020} define a cloud as destroyed when less than 10\% of the initial mass remains. Besides the different definitions of cloud destruction, the largest departure is likely their inclusion of conduction. In fact, they find that a cloud with (\Twind, \rcl, $n_{\rm wind}$, $v_{\rm wind}$) = ($10^7$\,K, 10\,pc, $10^{-2}$\,cm$^-3$, 100\,km/s) is destroyed with conduction included but survives when neglecting conduction (Li 2021, priv. comm.). For concerns regarding resolution see Appendix \ref{App:ResolutionConvergence}.


\subsubsection{Cloud Crushing: Cold Dusty Clouds}
\label{sec:DiscussionComparisonDust}
Previous work that specifically studied the survival of dust in the cloud crushing problem has fixated on the supernova remnant context \citep[Mach $\sim$ 10, cf.][]{Silvia2010,Silvia2012,Kirchschlager2019,Priestley2019,priestley2021revisiting,slavin2020dynamics}. We focus our comparison on the most similar works to our own, that of \citet{Silvia2010,Silvia2012}, who performed ENZO \citep[][]{bryan2014enzo} simulations treating dust as passive, velocity tracer particles perfectly coupled to the gas. They include radiative cooling but evolve their simulations only $\sim$10 \tcc\ since their cloud leaves the box in that time.

\citet{Silvia2010,Silvia2012} model dust as 4096 tracer particles, with each particle representing a population of grains with a distribution of sizes. They include thermal sputtering in post-processing and find efficient destruction of grains smaller than $0.1\,$mm, suggesting efficient destruction when inertial sputtering is efficient. They find the amount of dust that survives varies widely depending on the composition (and to a more minor extent on metallicity at $Z > 100 Z_{\odot}$; \citealt{Silvia2012}); single element grains have $> 80\%$ survival while Al$_{2}$O$_{3}$ is completely destroyed and MgSiO$_{3}$ has 14\% survival. The range of survival is in broad agreement with Figure~\ref{fig:DSF_TdustdeathBracket} where we observe a range of complete dust destruction to complete dust survival depending upon the sputtering time and destruction temperature.

Our work on the survival of dust complements recent work by \citet{Girichidis2021}, which studies the \textit{in-situ} formation of molecular gas. \citet{Girichidis2021} perform simulations of initially warm, atomic $\sim$\tenFour\,K clouds subjected to a hot $10^6$\,K wind with density contrasts 10$^{2-3}$, including H$_2$ formation catalyzed by dust grains. They assume a constant dust-gas ratio 0.01 (representative of the ISM), implicitly assuming perfect survival of dust and that dust remains coupled to their clouds. As we have shown, most dust may survive depending on the destruction model, motivating more detailed dust destruction modeling in future work. Their inclusion of effects we neglect, such as self-gravity of the gas, magnetic fields, a chemical network, and various heating terms limits our ability to compare our simulated clouds. In any case, they focus on the initial ($t \lesssim 10 t_{\rm cc}$) evolutionary stages of molecular clouds, whereas we study the conditions for cold (possibly molecular) cloud survival so again our work is complementary.
 
\subsection{Implications for Molecular Galactic Winds and the Circumgalactic Medium}
\label{sec:DiscussionGW}
Since the pioneering work of \citet{Spitzer1956}, the question of the cold gas content of the CGM has persisted. Recent absorption-line measurements by HST/COS detect massive reservoirs of cool 10$^{4}$\,K gas in the circumgalactic medium (\citealt{Werk2013}). The origins of such gas is difficult to ascertain as it may form via thermal instability (\citealt{Field1965}; \citealt{Sharma2012}; \citealt{McCourt2012}; \citealt{Voit2015}; \citealt{Ji2018}; \citealt{Butsky2020}; \citealt{falle2020thermal}; \citealt{kupilas2021interactions}) or be entrained in galactic outflows. While observational constraints of the warm, atomic phase are emerging, the cold, molecular component remains difficult to study. Our finding that cold (possibly molecular) gas is suggested to increase deep into the CGM (elaborated below) is intriguing, as such a growth may imply eventual star formation such as recently observed in galactic outflows (\citealt{maiolino2017star}; \citealt{gallagher2019widespread}). Such stars would possess highly radial orbits, developing the spheroidal component of galaxies and possibly evolving spiral into elliptical galaxies.

\subsubsection{Clouds in the Galactic Center}
\label{sec:DiscussionGC}
Recent 21-cm emission surveys \citep[][]{McClureGriffiths2013,DiTeodoro2018,lockman2020observation} have discovered a large sample of $\sim$200 neutral hydrogen clouds between $\sim$0.5-3.5 kpc from the Galactic midplane in the region commonly referred to as the Fermi bubbles \citep[][]{su2010giant}. Interestingly, molecular gas is observed to be spatially coincident with at least some of these warm clouds \citep[][]{Di2020,su2021molecular} and similarly for the Small Magellanic Cloud \citep{McClure2019,Di2019}. According to our model, all these clouds may have originally been cold (and molecular) and either some clouds' molecular content is below the sensitivity limit of existing observations or these clouds were originally in the regime of \med\ placing them in the ``warm survived'' regime. That is, they only possess warm, neutral gas now but if dust survives, the observed clouds will re-form molecular gas at large scale heights from the disk, deeper into the CGM.

\citet{DiTeodoro2018} observed most clouds to possess a characteristic size of $\sim$10s pc. Given their observed distance from the Galactic center, \citet{lockman2020observation} estimate the required survival times of the clouds to be $\gtrsim$4-10 Myr.  
These time-scales are rather perplexing in the classical cloud crushing picture because $t_{\rm cc}\sim 0.3 \,$Myr (using $v_w = 300\,{\rm km}\,{\rm s}^{-1}$, $\chi$ = \tenThree, \rcl\ = 10\,pc.) -- but quite sensible given our results. \med\ is the typical size compared to observations and exhibits a survival time of $\gg$ 10\,Myr\footnote{Our wind speed was 180 km/s, about two times lower than the best fit for \citet{DiTeodoro2018}. Observed clouds may be at higher overdensities or lower temperatures than simulated; therefore, caution should be used in comparing our results to that of \citet[][cf. \S \ref{sec:DiscussionCaveats}]{DiTeodoro2018}.}.

Our cold clouds exhibit mass growth of the \tenFour\,K phase very rapidly, potentially explaining the larger mass in neutral clouds observed at higher latitude by \citet{DiTeodoro2018} compared to the low latitude clouds of \citet{McClureGriffiths2013}. Note the larger volume sampled and higher sensitivity in the measurements of \citet{DiTeodoro2018} partly account for the mass increase, yet they find no evidence for mass loss with increasing latitude. More extensive observations at uniform sensitivity to determine the mass-dependence on latitude (similar to the work by \citealt{lockman2020observation} on the velocity-dependence) may place helpful constraints on cold cloud acceleration models.

\subsubsection{Extragalactic Multiphase Winds \& CGM}
\label{sec:DiscussionEG}
Galactic winds are ubiquitously detected throughout the observable Universe, and we now understand their vital role in galaxy evolution \citep[see reviews by][]{Rupke2018,Veilleux2005,Veilleux2020}. 
While the driving mechanism of these winds predominantly accelerates hot gas (e.g., supernovae and AGN; \citealt{Chevalier1985}) a characteristic feature of these winds is fast moving, cold gas. Thus, we now know that galactic winds are multiphase, i.e., characterized by co-spatial hot ($\gtrsim 10^6\,$K) and cold ($\lesssim 10^4\,$K) gas -- a picture that is also directly established by observations of nearby galaxies \citep[e.g.,][]{Heckman1995,Martin2005,strickland2009supernova}.

As eluded to in \S \ref{sec:intro}, how this much colder gas phase exists in such a hot, violent environment is an outstanding puzzle \citep[see][for a detailed discussion]{zhang2017entrainment} to which essentially two classes of solutions have been put forward in the literature: (i) accelerating the cold gas, or (ii) creating the cold phase out of the hot medium. The major obstacle for the former class of models has to overcome is the `entrainment problem'; that is, the fact that the acceleration time is shorter than the destruction time of the cold gas (in the case of ram-pressure acceleration by a factor of $\chi^{1/2}$, i.e., $\gtrsim 10$ for the temperatures quoted above). Thus far in the literature cooling \citep{Armillotta2016,Gronke2018}, magnetic fields \citep{Dursi2008,McCourt2015}, or `shielding effects' \citep{McCourt2018} have been suggested as potential solutions to this problem. 
For the second class of models in which the cold gas is forming from the hot medium, the cooling time of the (expanding) hot wind has to be shorter than other dynamical time-scales (notably $\sim r/v_{\rm wind}$) which is the case for a certain part of the parameter space \citep{Wang1994,Thompson2016,bustard2016versatile,Scannapieco2017,Schneider2018,Kempski2020,lochhaas2021characteristic}.

The fast accelerating field of detections of an even colder, molecular gas phase embedded in galactic winds \citep[e.g.,][]{Fischer2010,Cicone2014,Cicone2018} as well as of dust in the winds of the CGM \citep{Menard2010,PeekDUSTCIRCUMGALACTIC2015} offer an opportunity to differentiate between the mechanisms at play; in order to (re-)form efficiently, dust is a crucial ingredient \citep{Draine2011}. However, as dust gets destroyed in the hot medium (a) through shocks on a $\lesssim 1000\,$yr time-scale \citep{Dwek1996,Ferrara2016}; i.e., nearly instantaneously, and (b) through thermal sputtering on a time-scale of $\sim 5.5 (0.01\,{\rm cm}^{-3}/n) [(2\times 10^6 / T)^{2.5}+1]\,$Myr \citep{tsai1995interstellar,gjergo2018dust}. Combined, these effects make hot galactic winds a hostile environment for dust as was pointed out by \citet{Ferrara2016}. These authors simulated dust destruction in a hot, outflowing wind parcel and found survival times of only $\sim 10^4$ years and conclude that the detection of molecules in quasar outflows is difficult to explain theoretically.

The results of this work show that (a) the direct acceleration of cold ($\lesssim 1000\,$K and possibly molecular) gas by a hot wind is possible if \tmax/\tcc\ < 1 is fulfilled. This criterion corresponds to a gas cloud of size
\begin{equation}
    r_{\rm cl} \gtrsim r_{\rm cl,max} \equiv 50\,\mathrm{pc} \frac{T_{\rm cl,3}^{1/2}\Mach_{1.5}}{P_3 \Lambda_{-26}(T_{\rm max})}
    \label{eq:rccrit_max}
\end{equation} 
where $T_{\rm cl,3} \equiv $ (\Tcl/\tenThree\,K), \Mach$_{1.5} \equiv$ (\Mach/1.5), and $\Lambda_{-26}(T_{\rm max}) \equiv (\Lambda/10^{-26} {\rm erg}\,{\rm s}^{-1}\,{\rm cm}^3)$ is the cooling curve evaluated at $T_{\rm max}\sim 8000\,$K. Note that this equation is only valid when \Tcl\ < 8000\,K.

Furthermore, even for smaller clouds we show that (b) the `warm' $\sim 10^{4}$\,K gas can survive the ram pressure acceleration process if \tminmix/\tcc\ < 1, corresponding to a geometrical criterion of 
\begin{equation}
    r_{\rm cl} \gtrsim r_{\rm cl,minmix} \equiv 7\,\mathrm{pc} \frac{T_{\rm cl,3} \chi_3 \Mach_{1.5}}{P_3 \Lambda_{-21.6}(T_{\rm minmix})}
    \label{eq:rccrit_minmix}
\end{equation}
where $\chi_3 \equiv (\chi/10^3)$ and $\Lambda_{-21.6}(T_{\rm minmix}) \equiv (\Lambda/10^{-21.6} {\rm erg}\,{\rm s}^{-1}\,{\rm cm}^3)$ is the cooling curve evaluated at $T_{\rm minmix} \sim 2 \times 10^4\,$K. Note the above equation is accurate within a factor of two for 200\,K < \Tcl\ < 5000\,K. For \Tcl\ < 200\,K, $\Lambda(T_{\rm minmix})$ jumps sharply four orders of magnitude and thus one must include the actual value of $\Lambda(T)$. Thus, because embedded in this gas the sputtering time-scale of dust becomes very long (see above), it can survive, and allow the formation of molecules at larger distances.

These considerations suggest that detections of molecules and / or dust outflowing at distances $d \gtrsim \tau_{\rm sputter}/(4\, \mathrm{Myr}) v_{\rm wind}/(250\,\mathrm{km}\,\mathrm{s}^{-1})\,$kpc or a $\sim$ constant molecular abundance as a function of radius can be interpreted as signs of successful cold gas acceleration. Thus far, only a few of such cases are studied \citep[e.g.,][]{Walter2017} but future observations will enlarge the sample size and allow for a detailed mapping of the destruction and / or entrainment process as a function of distance for a range of galaxies.

Such observations could in principle detect also the second signpost of cold gas acceleration permitted and facilitated by cooling discussed in \S \ref{subsec:Sandman}, namely the very rapid acceleration due to an initial phase of mass loss followed by growth. Specifically, cold gas is detected moving `unnaturally' fast -- already at the hot gas velocity within distances of $d \lesssim v_{\rm wind} t_{\rm drag} \sim 50 \chi_3 r_{\rm cl}/(50\,\mathrm{pc})\,\mathrm{kpc}$ -- which would give direct evidence for an efficient mass / momentum transfer from the hot to the cold medium via cooling.


\subsection{Implications for Jellyfish Galaxies}
\label{sec:DiscussionJelly}
Galaxies undergoing extreme ram pressure stripping host multiphase tails extending $\sim$100\,kpc (e.g., D100 in the Coma Cluster, \citealt{jachym2017molecular}; cf. the GASP survey for additional examples, \citealt{poggianti2017gasp}). Recent multiwavelength measurements have detected signatures of molecular gas, dust, and star formation in a growing sample (cf. \citealt{moretti2018gasp}) of ``jellyfish'' galaxy tails, named in relation to the appearance of their star-forming tails, extending roughly 100 kpc from their galactic disks (\citealt{cortese2007strong}). Widespread star formation in ram-pressure stripped tails is enticing theoretically, as it could significantly contribute to the observed intracluster light.

However, the fact that molecular gas has been observed in these tails is theoretically puzzling since molecular gas is difficult to directly strip (\citealt{tonnesen2012star}). High-resolution CO measurements of three Virgo cluster galaxies by \citet{lee2017effect} indicate spatial coincidence of molecular gas with the stripped HI, but no clear sign of molecular gas stripping. Additional evidence suggests most of the molecular gas observed in jellyfish tails likely formed \textit{in-situ}: \citet{moretti2018gasp} observed four jellyfish galaxies with massive molecular tails (15-100\% of the stellar mass in the disk). Each galaxy had a similar amount of molecular gas in the disk to normal galaxies, suggesting inefficient stripping of disk molecular gas. Moreover, the molecular tail gas mass was comparable to the mass in neutral hydrogen that was stripped. 

Only in the most extreme cases of jellyfish galaxies is direct ram pressure stripping of molecular gas observed. ALMA observations of ESO137-001 close to the center ($\sim$250 kpc) of the Norma cluster \citep{jachym2019alma} detect $\sim$kpc sized molecular cloud complexes, likely dynamically stripped while the spiral arm was unwound (ESO137-001's extant HI disk is only $\sim$1 kpc). Even in this case, the observations are surprising since the cloud crushing time of this molecular cloud complex is relatively short compared to the time from which it was expected to be stripped.

Our work elucidates these observations of molecular gas in jellyfish tails. We naturally expect $\sim$kpc molecular clouds to survive intact. Moreover, we have shown that dust can survive the acceleration process for \tenThree\,K clouds and hence stripped dusty, atomic gas can condense to form molecules. That dust survives ram pressure stripping is consistent with imaging from \textit{HST} \citep[][]{elmegreen2000dust,cramer2019spectacular} and \textit{Herschel} \citep[][]{cortese2010herschel}. Moreover, in our model the cold gas mass can naturally exceed the original stripped mass by accreting ICM material. This picture is validated by the detection in JO201 of a metallicity gradient from the disk towards the tail, suggesting disk material mixes with the ICM \citep[][]{bellhouse2019gasp}.

Further validation of our model explaining molecular gas in jellyfish tails is suggested by combining our results with the work of \citet{Tonnesen2021}. The variety of cloud sizes and densities found to be stripped from galactic disks in \citet{Tonnesen2021} suggests some clouds will be similar to our \lrg\ and \med\ fiducial cloud runs. Since we find that abundant dust can survive, molecules can form, leading to subsequent star formation as observed. However, more detailed comparisons will require simulations including a chemical network to directly identify the evolution of molecular gas in our simulations, as well as density contrasts of $\chi = 10^{4-5}$, which we leave for future work. Similarly, including magnetic fields and self-gravity in future work may help constrain star formation in jellyfish tails \citep[][]{muller2021highly}.


\subsection{Caveats}
\label{sec:DiscussionCaveats}
We neglect a number of potentially important physical effects that may modify the results of this work, namely viscosity \citep{Li2020,Jennings2020}, conduction (\citealt{bruggen2016launching}; \citealt{Armillotta2017}; \citealt{Li2020}), external turbulence \citep[][]{vijayan2018extraplanar,banda2018filament,banda2019dynamics,Tonnesen2019,Schneider2020}, magnetic fields \citep{Gronnow2018,Gronke2020Cloudy}, and cosmic rays \citep{Wiener2017,Wiener2019,bruggen2020launching,Bustard2021}. 

While conduction can play a role in the shape of the cold gas \citep{bruggen2016launching}, and certainly does affect observables \citep{Tan2021b}, detailed simulations of turbulent radiative mixing layers have shown that the mass transfer between the phases is not affected by conduction in the `strong cooling' regime \citep{Ji2018,Fielding2020,tan2021radiative}. This is because turbulent diffusion dominates over thermal conduction, and the mass transfer rate is set by mixing on large scales and not by the microscopic (thermal or numerical) diffusion between the gas phases. This also implies that only the large scales (in our case $\sim r_{\rm cl}$) need to be resolved in order to obtain a converged mass growth rate \citep{tan2021radiative}.

We expect non-thermal components like cosmic rays and magnetic fields to affect our results. Prior studies focusing on $T_{\rm cl}\sim 10^4\,$K have in particular found that magnetic fields can suppress mixing \citep[e.g.,][in a plane-parallel setup]{Ji2018}; thus, we expect the mass growth rates to potentially change. Interestingly, while `magnetic draping' has been invoked to shield the cold gas, for larger overdensitites this effect becomes subdominant and for $\chi \gtrsim 100$, i.e., the ones considered in this study, it is not sufficient to make the gas survive until entrainment (\citealp{Gronke2020Cloudy,cottle2020launching}). We therefore do not expect our survival criteria to change dramatically with the inclusion of magnetic fields.

Similarly with cosmic rays: while their non-thermal support would decrease the overdensities of our blobs \citep[e.g.,][]{Butsky2020}, and thus allow, for instance, for faster entrainment, this would only affect $\chi$ and hence the time-scales considered by at maximum a factor of a few. 

We considered only an (initially) laminar wind impinging the cloud from one direction whereas in reality galactic winds have a non-radial component \citep{vijayan2018extraplanar,Schneider2020}. This turbulence will affect the cold gas (and the cooling flow, cf. \citealt{dutta2021cooling}) and a detailed study of this effect (building off \citealt{gronke2021survival}) is an interesting avenue for future work. However, we note that the non-radial component is subdominant over the shock caused by the wind part and is, thus, arguably less responsible for the destruction of the cold gas \citep[see, e.g., figure 21 in ][showing the non-radial component to be $\mathcal{M}\lesssim 0.1$]{Schneider2020}. This is consistent with the findings of \citet{Li2020} who concluded that turbulence would not affect their results significantly.

In this work, we focus our analysis on \Tcl = 10$^3$\,K clouds. Note that most previous work uses \Tcl = 10$^4$\,K which is an equilibrium point between radiative cooling and UV photoionization heating. However, this is a density-dependent statement and sufficiently dense clouds cool past this equilibrium to lower temperatures -- potentially very low temperatures of a few-tens\,K as molecular clouds. It is computationally intractable for us to perform simulations while maintaining the hot phase at 10$^6$\,K as the density contrast becomes quite large. Instead, we vary the initial cloud temperature and here focus on \Tcl = \tenThree\,K clouds to better compare with previous work including a 10$^6$\,K hot phase. Future work will simulate larger density contrasts.

In this work, we used the fixed effective cooling curve introduced in \S \ref{subsec:NumericalSetup}. In particular, we neglected metallicity, heating, and ionization effects which introduce additional dependencies on $\Lambda$. We note, however, that (a) the here derived criteria are general and can be adapted to a changed cooling curve, and (b) that our survival criteria imply column densities of $N\gtrsim r_{\rm cl,minmix} n\sim 10^{19}\,{\rm cm}^{-2}$ (since $r_{\rm cl,minmix}\propto 1/n$; \citealp{Gronke2020Cloudy}), i.e., the surviving objects are self-shielded. A spatially varying cooling curve, e.g., due to mixing of different metallicity gas or ionization, might have interesting effects for the mixing process and is an interesting avenue for future work.

Since we study the early evolution of relatively diffuse cold clouds, we neglect self-gravity. However, as the cold clouds grow in mass they may exceed the Jeans mass, become self-gravitating, and form stars (cf. \S~\ref{sec:DiscussionJelly}). We plan to include self-gravity in future work to determine the propensity of molecular galactic outflows to become starforming (as recently observed, e.g., \citealt{gallagher2019widespread}) and to study the observed star formation in jellyfish tails \citep[][]{vulcani2018enhanced}.

Perhaps of most concern to the conclusions of our work is the neglect of detailed dust modeling. We present our results in a general fashion, that is, as a function of destruction temperature and sputtering time (cf.~\S\ref{subsec:DustSurvival}) to allow for more more sophisticated dust models to be mapped onto them. We expect our chosen parameter range to bracket the more detailed dust models. 

For instance, as \citet{Kirchschlager2019} show thermal sputtering alone destroys only 20\% of dust in their model, whereas in combination with inertial sputtering nearly all dust is destroyed (cf. their figure 22). Studies of AGN heating of elliptical galaxies find near instantaneous and complete thermal sputtering when dust encounters the hot $\sim$10$^7$\,K ISM of elliptical galaxies \citep[][]{valentini2015agn}. Therefore, our treatment of instant dust destruction when dust encounters the hot phase roughly approximates the effect of efficient inertial sputtering shifting the dust distribution to small sizes followed by efficient thermal sputtering in the hot phase (that is, t$_{\rm sputter} \rightarrow 0$), or, dust encountering an unmodeled 10$^7$\,K phase. Allowing longer thermal sputtering times may be seen as decreasing the efficiency of inertial sputtering since in such a case larger grains will be present. Similarly, considering a range of destruction temperatures allows one to consider the survival of more volatile/robust grain species. We hope to explore these effect and perform detailed dust modeling in future work. 

An additional concern is that we model dust as passive tracers perfectly coupled to the gas, neglecting processes such as the resonant drag instability \citep[][]{squire2018resonant}. Including the drag between gas and dust may result in more dust entering the hot medium, reducing the dust survival. We hope to explore this effect in future work.

We furthermore hope to extend this work to higher Mach numbers as well as overdensitites, and more realistic wind geometries that we omitted in this work.

\section{Conclusions}
\label{sec:Conclusions}
We perform simulations of cold clouds subjected to a hot wind. In our simulations we find that generally dust and cold gas can survive ram pressure acceleration, if $\tmax/\tcc < 1$. However, we also show that the survival of dust depends on the time it can withstand a hot surrounding, as no gas stays cold the entire time but instead cold gas mixes with the hot wind during the acceleration process and cools back to lower temperatures. Our simple model is consistent with previous work considering dust destruction in supernova remnant shocks \citep[][]{Silvia2010,Priestley2019,Kirchschlager2019,slavin2020dynamics}. Moreover, we find the \tmixCool/\tcc\ $<$ 1 condition of \citet{Gronke2018} determines whether $\sim$\tenFour\,K clouds survive or are ablated by the hot wind, in agreement with recent studies (\citealt{Abruzzo2021}; \citealt{Kanjilal2021}). 

A novel aspect of our work is the consideration of dust destruction in cold clouds in a galactic wind context. We model dust as one million passive velocity tracer particles, assuming perfect coupling to the gas. In post-processing we determine the ability of dust to survive the acceleration process with a simple model, depending on the time ($t_{\rm sputter}$) dust is in contact with hot gas (T $>$ \Tdest). Our specific conclusions are as follows:

\begin{enumerate}
    \item \textit{Survival Criteria for Cold Gas}. We discover a three-path solution for the evolution of cold ($\sim$\tenThree\,K) clouds subjected to a hot ($10^6$\,K) wind. Radiative cooling becomes inefficient below the temperature at which hydrogen becomes predominantly neutral, forming a local maximum in the cooling time curve \tmax\ $\sim$8000\,K. By considering a variety of initial cloud sizes for \Tcl\ = \tenThree\,K clouds we find survival for cold gas requires \tmax/\tcc\ $<$ 1, yet warm ($\sim$\tenFour\,K) gas can persist so long as cooling is more efficient than mixing between the warm cloud phase and the hot wind; that is, \tminmix/\tcc\ $<$ 1. These two possible evolutionary paths merge and recover to the survival condition of \citet{Gronke2018} for \Tcl\ $\geq$ \tenFour\,K as the cooling time curve is monotonic for T $\geq 2 \times$\tenFour\,K.
    
    \item \textit{Growth vs. Mass Loss Condition for Surviving Clouds}. Even when cold gas survives there is typically a period of mass loss before the cold mass eventually grows (up to 10 times the initial mass). We determined \tmax/\tmixCC\ $<$ 1 is the condition for mass growth. That is, efficient gas cooling below the peak of the cooling time curve \tmax\ must exceed the instantaneous mixing rate, set by the eddy turnover time \tmixCC (which can initially be much smaller than the cloud crushing time, but grows long as the turbulent velocity drops as the cloud becomes entrained). This result might explain similar growth profiles observed previously in the literature \citep[e.g.,][]{Sparre2020,Kanjilal2021,Abruzzo2021}.
    
    \item \textit{Fast Entrainment of Cold Gas}. We find cold clouds are entrained much more rapidly than warm clouds at the same initial overdensity $\chi = 10^3$ and similar cooling efficiency; this is faster than the acceleration time $t_{\rm drag}$ theoretically expected from our initial conditions. Our simulated cold clouds are rapidly cocooned by a warm \tenFour\,K phase which efficiently exchanges momentum with the hot gas, allowing for the cold phase to be accelerated $\sim$3 times faster than direct mixing between the cold and hot phases. This helps to explain observations of cold gas rapidly outflowing from galaxies close to the expected launching radius (cf. \citealp{Veilleux2020}).
    
    \item \textit{Survival of Dust}. In simulations of cold clouds that survive the acceleration in a hot wind, we find that the survival of dust is possible but depends sensitively on the destruction temperature and cloud size. For \Tdest\ $\geq$ \tenFour\,K, unless dust can inhabit warm-hot gas for $>$ 30 Myr in the case of large clouds we find $>$ 90\% of dust will be destroyed. Destruction temperatures slightly above \tenFour\,K allow nearly all dust to survive in $\sim$100\,pc clouds, whereas $\sim$10\,pc clouds require dust to persist in hot gas at least a few Myr. The rather sharp transition between the vast majority of dust surviving versus destroyed may allow observations of dust in galactic winds and ram pressure stripped tails to constrain models of dust destruction.
\end{enumerate}
While these results help to understand cold gas dynamics -- and the survival of dust -- in galactic winds, many open questions remain such as the inclusion of more relevant physics, the expansion of the considered parameter space, and boundary conditions closer to reality. We plan to pursue these in future work.

\section*{Acknowledgements}
We thank the referee for a helpful review. RJF gratefully acknowledges Mateusz Ruszkowski, Paco Holguin, Chaoran Wang, Zhijie Qu, Stephanie Tonnesen, Chad Bustard, Alankar Dutta, Prateek Sharma, Zhihui Li, Martin Sparre, and Greg Bryan for helpful discussions. 
RJF thanks the Kavli Institute for Theoretical Physics for hospitality during the start of this project as part of the Graduate Fellowship Program. This research was supported in part by the National Science Foundation under Grant No. NSF PHY-1748958, and XSEDE grant TG-AST180036. 
MG was supported by NASA through the NASA Hubble Fellowship grant HST-HF2-51409 and acknowledges support from HST grants HST-GO-15643.017-A, and HST-AR15039.003-A. 
This project utilized the visualization and data analysis package \texttt{yt} \citep[][]{Turk2011}; we are grateful to the yt community for their support.

\section*{Data Availability}

Data related to this work will be shared on reasonable request to the corresponding author.



\bibliographystyle{mnras}
\bibliography{paper_dust_survival} 



\appendix

\section{Resolution Convergence}
\label{App:ResolutionConvergence}

\begin{figure}
  \begin{center}
    \leavevmode
    \includegraphics[width=0.8 \linewidth]{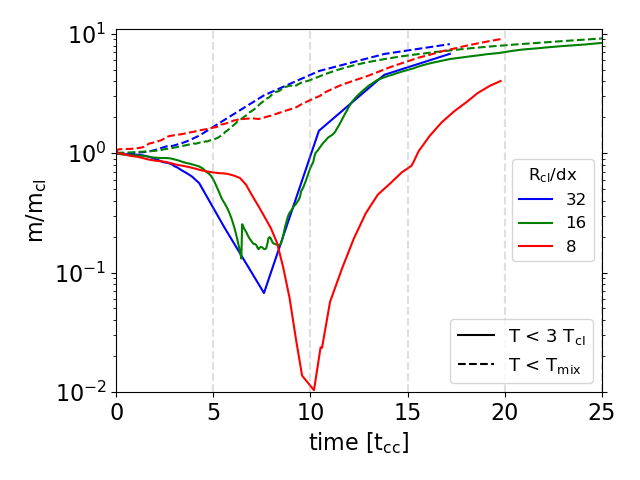}
\caption[]{Mass growth converges with increasing resolution. We are showing here the \Tcl\ = $10^3$\,K, \rcl\ = 100\,pc (\lrg) run. The red, green, and blue curves refer to 8, 16, and 32 cells per cloud radius resolution respectively. Solid (dashed) curves indicate the mass in the cold (warm and cold) phase(s).}
\label{fig:MassConvergenceResolution}
\end{center}
\end{figure}

We performed simulations with the same setup as \lrg\ at two times coarser and two times higher resolution (8 and 32 cells per cloud radius) to test convergence. We find larger initial destruction of cold gas and a longer period before growth of the low resolution run (8 cells per cloud radius), whereas the fiducial (16 cells per cloud radius) and high resolution (32 cells per cloud radius) evolve remarkably similarly. Note that both low and high resolution simulations were performed with the same box dimensions as the fiducial run (see Table \ref{tab:ListOfRuns}).

\section{Dust Survival Fraction Convergence}
\label{App:DustSurvivalFractionConvergence}
\begin{figure}
  \begin{center}
    \leavevmode
    \includegraphics[width=0.8 \linewidth]{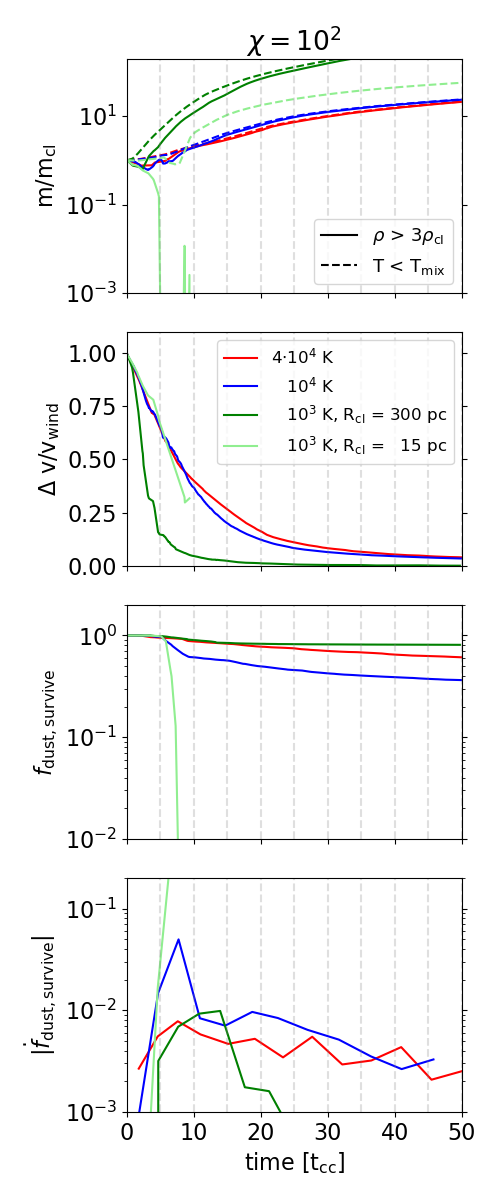}
\caption[]{Dust survival fraction converges as shear velocity asymptotes to zero. The bottom row shows the logarithmic slope of the dust survival fraction, demonstrating convergence (with simulation time) of the destruction rate.}
\label{fig:MassShearDSFChi100}
\end{center}
\end{figure}

To investigate the convergence of the dust survival fraction, we perform simulations at $\chi = 10^2$ since the dynamical time is shorter and hence the simulations arrive at a saturated state more readily than at higher overdensities. In Figure \ref{fig:MassShearDSFChi100} we see that the dust survival fraction (third row) has $\sim$saturated as the shear velocity approaches zero, as evidenced by the time derivative of the survival fraction (fourth row) dropping below 10$^{-2}$. Also note that the mass growth has similarly saturated; dust survival and mass growth are anticorrelated at late times since they both involve mixing of cloud and wind material.

\section{Dust Initial Conditions}
\label{sec:DustInitialConditions}

\begin{figure}
  \begin{center}
    \leavevmode
    \includegraphics[width=0.99 \linewidth]{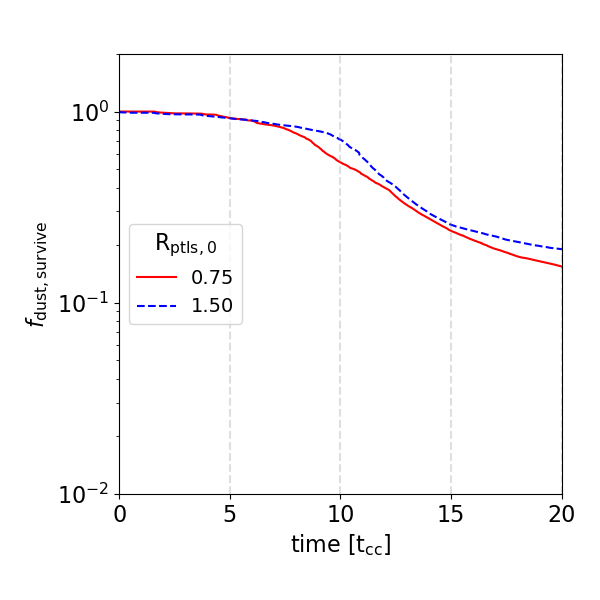}
\caption[]{Dust survival fraction does not strongly depend on initial placement of dust.}
\label{fig:DustInitialPlacement}
\end{center}
\end{figure}

To determine the impact of our placement of dust within a subvolume of the initial cloud, we performed one simulation in which dust was initialized randomly uniformly within 1.5 \rcl\ from cloud center. We neglected dust initially placed outside the cloud from the dust survival fraction analysis, reducing the number of dust particles involved in the analysis to $\sim$300,000. From Figure \ref{fig:DustInitialPlacement} one can see the initial placement of dust only has a $\sim$10\% impact on the survival fraction.


\bsp	
\label{lastpage}
\end{document}